\documentclass[a4paper,fleqn,usenatbib]{mnras}

\usepackage{newtxtext,newtxmath}

\usepackage[T1]{fontenc}
\usepackage{ae,aecompl}

\usepackage{graphicx}	
\usepackage{amsmath}	
\usepackage{threeparttable,adjustbox,array,multirow,makecell}
\usepackage{breqn}

\newenvironment{altfont}{\fontfamily{lmtt}\selectfont}{\par}
\DeclareTextFontCommand{\textaltfont}{\altfont}
\defcitealias{C20}{Paper I}
\defcitealias{C21}{Paper II}
\defcitealias{vandermarelMagellanicCloudStructure2001}{vdM01}
\defcitealias{vandermarelNewUnderstandingLarge2002}{vdM02}
\title[MagES: LMC Outskirts]{The Magellanic Edges Survey III. Kinematics of the disturbed LMC outskirts}

\author[L. R. Cullinane et al.]{L. R. Cullinane$^{1}$\thanks{E-mail: lara.cullinane@anu.edu.au (LRC)},
A. D. Mackey$^{1}$,
G. S. Da Costa$^{1}$,
D. Erkal$^{2}$,
S. E. Koposov$^{3,4}$, \newauthor
V. Belokurov$^{4}$
\\
$^{1}$Research School of Astronomy and Astrophysics, Australian National University, Canberra, ACT 2611, Australia\\
$^{2}$Department of Physics, University of Surrey, Guildford GU2 7XH, UK \\
$^{3}$Institute for Astronomy, University of Edinburgh, Royal Observatory, Blackford Hill, Edinburgh EH9 3HJ, UK\\
$^{4}$Institute of Astronomy, University of Cambridge, Madingley Road, Cambridge CB3 0HA, UK\\
}
\date{Accepted XXX. Received YYY; in original form ZZZ}

\pubyear{2021}

\begin{document}
\label{firstpage}
\pagerange{\pageref{firstpage}--\pageref{lastpage}}
\maketitle

\begin{abstract}
We explore the structural and kinematic properties of the outskirts of the Large Magellanic Cloud (LMC) using data from the Magellanic Edges Survey (MagES) and Gaia EDR3. Even at large galactocentric radii (8$^\circ$<$R$<11$^\circ$), we find the north-eastern LMC disk is relatively unperturbed: its kinematics are consistent with a disk of inclination \textasciitilde36.5$^\circ$ and line-of-nodes position angle \textasciitilde145$^\circ$ east of north. In contrast, fields at similar radii in the southern and western disk are significantly perturbed from equilibrium, with non-zero radial and vertical velocities, and distances significantly in front of the disk plane implied by our north-eastern fields. We compare our observations to simple dynamical models of the Magellanic/Milky Way system which describe the LMC as a collection of tracer particles within a rigid potential, and the Small Magellanic Cloud (SMC) as a rigid Hernquist potential. A possible SMC crossing of the LMC disk plane \textasciitilde400~Myr ago, in combination with the LMC’s infall to the Milky Way potential, can qualitatively explain many of the perturbations in the outer disk. Additionally, we find the claw-like and arm-like structures south of the LMC have similar metallicities to the outer LMC disk ([Fe/H]\textasciitilde$-$1), and are likely comprised of perturbed LMC disk material. The claw-like substructure is particularly disturbed, with out-of-plane velocities >60~km~s$^{-1}$ and apparent counter-rotation relative to the LMC’s disk motion. More detailed $N$-body models are necessary to elucidate the origin of these southern features, potentially requiring repeated interactions with the SMC prior to \textasciitilde1~Gyr ago.
\end{abstract}




\section{Introduction}
At respective distances of \textasciitilde50 and \textasciitilde60~kpc \citep{pietrzynskiDistanceLargeMagellanic2019,graczykDistanceDeterminationSmall2020a}, the Large and Small Magellanic Clouds (LMC/SMC) are the closest pair of interacting Milky Way (MW) dwarf satellites. This affords us the unique opportunity to study in detail the effects of tidal interactions on both the dynamics and star-formation history of the Magellanic system. As the Clouds are thought to be on their first infall into the Milky Way potential \citep{beslaAreMagellanicClouds2007,kallivayalilThirdEpochMagellanicCloud2013}, the plethora of unusual features observed in the Clouds are likely the result of repeated interactions between the two galaxies themselves, with the Milky Way affecting their morphology and dynamics only comparatively recently. These features include the extensive Magellanic Stream of HI gas \citep[e.g.][]{mathewsonMagellanicStream1974,nideverOriginMagellanicStream2008a}, the irregular morphology and dynamics of the SMC \citep[e.g.][and many others]{hatzidimitriouStellarPopulationsLargescale1989,harrisSpectroscopicSurveyRed2006a,dobbieRedGiantsSmall2014b,ripepiVMCSurveyXXV2017,deleoRevealingTidalScars2020a}, and the tilted, off-centre stellar bar and single spiral arm in the LMC \citep{devaucouleursStructureDynamicsBarred1972,vandermarelMagellanicCloudStructure2001}.

While the less massive SMC is the more heavily distorted of the two galaxies, the more massive LMC has not escaped from these interactions unscathed. The morphology of the LMC can be broadly described as that of an inclined disk, but it also displays significant deviations from simple ordered rotation in the disk plane. In addition to the unusual stellar bar and spiral arm, both of which are predominantly comprised of younger stars \citep[e.g.][]{elyoussoufiVMCSurveyXXXIV2019}, older stellar populations also show evidence of perturbation, including multiple warps \citep{choiSMASHingLMCTidally2018,olsenWarpLargeMagellanic2002}, ring-like overdensities \citep{kunkelDynamicsLargeMagellanic1997,choiSMASHingLMCMapping2018}, and offsets between the observed photometric and kinematic centres \citep[see e.g. Table 1 of][for a review]{wanSkyMapperViewLarge2020}, including differences in kinematic centres for different tracer populations. Further, deep photometric studies of the Magellanic periphery \citep[e.g.][]{mackey10KpcStellar2016,mackeySubstructuresTidalDistortions2018,pieresStellarOverdensityAssociated2017a}, in combination with multi-dimensional phase-space information from Gaia \citep[e.g.][]{belokurovCloudsStreamsBridges2017,belokurovCloudsArms2019,gaiacollaborationGaiaEarlyData2021a}, have revealed an abundance of stellar substructure surrounding the Clouds. This includes claw-like structures in the southern LMC outskirts and an apparent truncation in the western edge of the LMC disk \citep{mackeySubstructuresTidalDistortions2018}, diffuse structures to the east of the LMC \citep{youssoufiStellarSubstructuresPeriphery2021}, a diffuse overdensity to the northwest of the SMC \citep{pieresStellarOverdensityAssociated2017a}, a long, thin feature which appears to wrap around the southern LMC, stretching between the eastern outskirts of the SMC and the eastern LMC disk \citep{belokurovCloudsArms2019}, and a \textasciitilde23$^\circ$ long arm-like feature to the north of the LMC \citep{mackey10KpcStellar2016,belokurovCloudsArms2019}.

In order to constrain the complex interactions which produce these features, kinematic data are critical. Recently, in \citet{C21} we analysed the origin of the LMC’s northern arm using 3D kinematics from the Magellanic Edges Survey \citep[MagES:][]{C20} -- a spectroscopic survey of red clump (RC) and red giant branch (RGB) stars using the 2dF/AAOmega instrument \citep{lewisAngloAustralianObservatory2dF2002,sharpPerformanceAAOmegaAAT2006} on the 3.9~m Anglo-Australian Telescope (AAT) at Siding Spring Observatory -- in conjunction with Gaia astrometry. We found the arm is likely perturbed by a combination of the LMC’s current infall to the Milky Way, and historical interactions with the SMC prior to \textasciitilde1~Gyr ago. This represents clear kinematic evidence that interactions between the Clouds, prior to the recent pericentric passage of the SMC \textasciitilde150~Myr ago \citep{zivickProperMotionField2018b}, could have produced measurable effects on the present-day structure of the Clouds. 

In this paper, we extend our MagES analysis to provide a comprehensive overview of the structural and kinematic properties of the outer LMC disk and surrounding stellar substructures. A summary of the data is presented in Section~\ref{sec:data}. Section~\ref{sec:fitdisk} describes how red clump photometry is utilised to derive a distance scale to the observed structures, and Section~\ref{sec:pops} examines the stellar population properties in the outer LMC that allow for such an analysis. The resultant structural and kinematic properties of the LMC outskirts are presented in Section~\ref{sec:3dkins}. We discuss the properties of each kinematically distinct region, and the implications for interactions between the Clouds and with the Milky Way, in Section~\ref{sec:analysis}. Section~\ref{sec:concs} summarises our conclusions. 

\section{Data}\label{sec:data}
\subsection{Observations}\label{sec:obs}
This paper is based on the analysis of 18 MagES fields targeting the outskirts of the LMC. Ten fields are located in the outer disk between LMC galactocentric radii of 8.5$^\circ$<$R$<11$^\circ$, five trace the long arm-like southern substructure discussed in \citet{belokurovCloudsArms2019}, and three are located on the claw-like southern substructures discussed in \citet{mackeySubstructuresTidalDistortions2018}. 

\begin{figure*}
	\includegraphics[height=12cm]{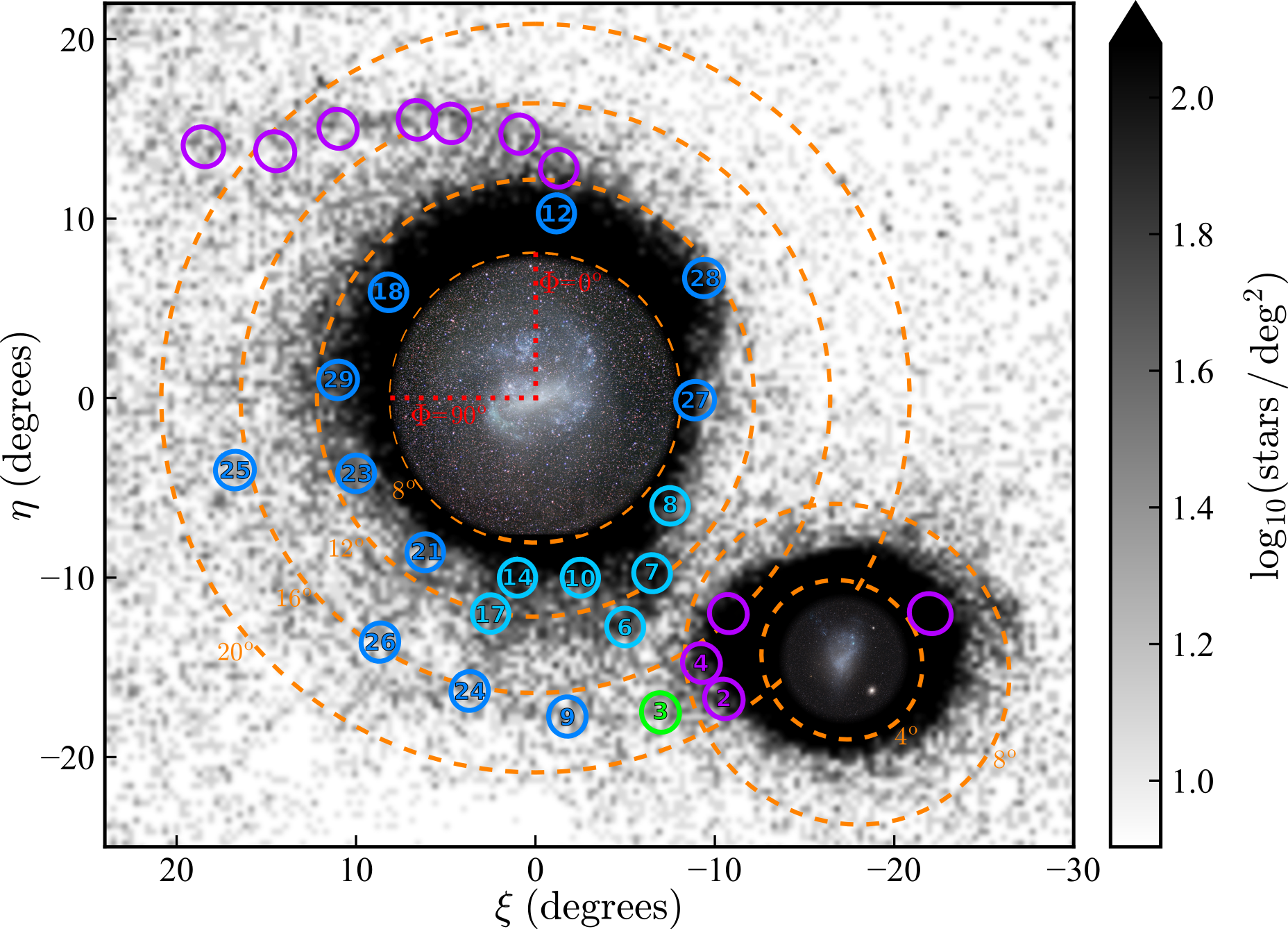}
	\caption{Location of observed MagES fields across the Magellanic Periphery. Blue fields are those analysed in this paper, with light and dark blue representing “M” and “G” fields respectively (see Section~\ref{sec:obs} for further detail on these classifications). Purple fields are MagES fields not analysed in this paper, and the green field is field 3, which is initially analysed but is found to be predominantly associated with the SMC, not the LMC. Any field discussed in the text is numbered. The background image shows the density of Magellanic red clump and red giant stars per square degree, selected from Gaia DR2 (the target catalogue from which most MagES stars are drawn) as per \protect\cite{belokurovCloudsArms2019}. On this map, north is up and east is to the left; ($\eta, \xi$) are coordinates in a tangent-plane projection centred on the LMC ($\alpha_0=82.25^{\circ}$, $\delta_0=-69.5^{\circ}$). Orange dashed circles mark angular separations of $8^\circ$, $12^{\circ}$, $16^{\circ}$ and $20^{\circ}$ from the LMC centre and $4^\circ$, $8^\circ$ from the SMC centre. Red dashed lines indicate position angles ($\Phi$) of 0 and 90 degrees east of north, for reference.}
	\label{fig:map}
\end{figure*}

The survey overview for MagES is presented in \citet[][henceforth referred to as Paper I]{C20}, and contains details of the target selection procedures, observation characteristics, and data reduction pipeline. Here we briefly summarise salient details for the fields used in this paper. As MagES is an ongoing survey, we note that three additional fields have been observed since the release of \citetalias{C20}: fields 27-29. These were observed on December 18 and 19 2020, with exposure times of 10800~s, 10055~s, and 10800~s respectively. We show details of these fields, in addition to existing MagES fields discussed in this paper, in Table~\ref{tab:fields}. The positions of all MagES fields are presented in Fig.~\ref{fig:map}, with the fields analysed in this paper indicated in blue. 

\begin{table*}
	\caption{MagES fields in the LMC outskirts analysed in this paper. Columns give the field number and classification as described in \protect\citetalias{C20}; location of the field centre as RA($\alpha$), DEC($\delta$) in J2000.0; on-sky distance of the field from the centre of the LMC ($R_{\text{LMC}}$) and position angle ($\Phi$) measured east of north, number of likely Magellanic stars per field, and aggregate kinematic parameters calculated as in \S\ref{sec:kindata}. The Magellanic population in field 3 is dominated by SMC stars, whilst all other fields are dominated by LMC stars.}
	\label{tab:fields}
	\begin{adjustbox}{max width=\textwidth}
	\begin{tabular}{lllrrrllllrl}
		\hline
		\multicolumn{1}{>{\centering\arraybackslash}m{1.5cm}}{Field (Class)} & \multicolumn{1}{c}{RA} & \multicolumn{1}{c}{DEC} & \multicolumn{1}{c}{$R_{\text{LMC}}$ ($^\circ$)} & \multicolumn{1}{c}{$\Phi$ ($^\circ$)} & \multicolumn{1}{>{\centering\arraybackslash}p{1.4cm}}{$N_{\text{Magellanic}}$ ($P_i\geq50\%$)} & \multicolumn{1}{>{\centering\arraybackslash}p{1.2cm}}{$V_{\text{LOS}}$ \newline(km~s$^{-1}$)} & \multicolumn{1}{>{\centering\arraybackslash}p{1.2cm}}{$\sigma_{\text{LOS}}$\newline (km~s$^{-1}$)} & \multicolumn{1}{>{\centering\arraybackslash}p{1.2cm}}{$\mu_\alpha$\newline (mas~yr$^{-1}$)} & \multicolumn{1}{>{\centering\arraybackslash}p{1.2cm}}{$\sigma_\alpha$\newline (mas~yr$^{-1}$)} & \multicolumn{1}{>{\centering\arraybackslash}p{1.2cm}}{$\mu_\delta$\newline (mas~yr$^{-1}$)} & \multicolumn{1}{>{\centering\arraybackslash}p{1.2cm}}{$\sigma_\delta$\newline (mas~yr$^{-1}$)} \\ \hline
		3 (G) & 01   20 00  & -82   30 00  & 18.2 & 201.7 & 65 & $185.4\pm4.1$ & $31.4\pm3 .2$ & $1.41\pm0.07$ & $0.48\pm0.05$ & $-1.37\pm0.04$ & $0.27\pm0.05$ \\
		6 (M) & 03   22 33  & -80   40 55  & 13.1 & 201.2 & 29 & $174.5\pm3.5$ & $19.4\pm3.1$ & $1.78\pm0.06$ & $0.35\pm0.06$ & $-0.52\pm0.04$ & $0.11\pm0.06$ \\
		7 (M) & 03   26 04  & -77   26 18  & 11.0 & 213.5 & 64 & $167.6\pm3.3$ & $34.1\pm2.5$ & $1.61\pm0.04$ & $0.30\pm0.03$ & $-0.61\pm0.03$ & $0.16\pm0.04$ \\
		8 (M) & 03   39 15  & -73   43 48  & 8.8 & 231.3 & 97 & $197.5\pm2.6$ & $25.0\pm2.0$ & $1.77\pm0.03$ & $0.15\pm0.03$ & $-0.60\pm0.02$ & $0.14\pm0.04$ \\
		9 (G) & 03   40 00  & -86   17 13 & 17.1 & 185.6 & 52 & $198.1\pm3.4$ & $18.3\pm2.3$ & $2.50\pm0.04$ & $0.21\pm0.03$ & $-0.13\pm0.05$ & $0.30\pm0.04$ \\
		10 (M) & 04   36 23  & -79   07 17  & 9.9 & 193.5 & 77 & $211.2\pm2.4$ & $20.3\pm1.9$ & $2.02\pm0.03$ & $0.21\pm0.04$ & $-0.01\pm0.03$ & $0.21\pm0.03$ \\
		12 (G) & 05   20 00  & -59   18 00  & 10.3 & 355.4 & 284 & $287.1\pm1.5$ & $24.8\pm1.1$ & $1.78\pm0.01$ & $0.12\pm0.01$ & $0.20\pm0.01$ & $0.19\pm0.01$ \\
		14 (M) & 05   50 22  & -79   21 18  & 10.0 & 173.6 & 85 & $244.8\pm2.8$ & $25.2\pm2.1$ & $1.97\pm0.03$ & $0.21\pm0.03$ & $0.80\pm0.03$ & $0.21\pm0.04$ \\
		17 (M) & 06   32 16  & -80   59 36  & 12.2 & 167.7 & 108 & $239.6\pm2.7$ & $26.0\pm2.1$ & $1.84\pm0.02$ & $0.17\pm0.02$ & $1.28\pm0.03$ & $0.30\pm0.03$ \\
		18 (G) & 06   40 00  & -62   30 00  & 10.7 & 55.4 & 299 & $324.5\pm1.2$ & $20.3\pm0.9$ & $1.49\pm0.01$ & $0.11\pm0.01$ & $1.00\pm0.01$ & $0.12\pm0.01$ \\
		21 (G) & 07   17 12  & -76   36 00  & 10.9 & 143.9 & 149 & $275.9\pm1.9$ & $21.6\pm1.5$ & $1.40\pm0.02$ & $0.17\pm0.02$ & $1.57\pm0.03$ & $0.26\pm0.02$ \\
		23 (G) & 07   36 00  & -71   00 00  & 11.4 & 112.7 & 127 & $302.2\pm2.9$ & $30.2\pm2.0$ & $1.11\pm0.02$ & $0.13\pm0.02$ & $1.64\pm0.02$ & $0.17\pm0.02$ \\
		24 (G) & 07   58 48  & -84   12 00  & 16.4 & 167.1 & 56 & $225.1\pm2.1$ & $15.1\pm1.7$ & $1.09\pm0.04$ & $0.21\pm0.03$ & $2.23\pm0.03$ & $0.16\pm0.03$ \\
		25 (G) & 08   32 00  & -67   00 00  & 17.5 & 103.7 & 37 & $344.3\pm2.5$ & $13.7\pm1.9$ & $0.75\pm0.02$ & $0.05\pm0.02$ & $1.97\pm0.04$ & $0.16\pm0.03$ \\
		26 (G) & 08   48 00  & -79   00 00  & 16.1 & 147.3 & 37 & $258.4\pm3.2$ & $18.5\pm2.3$ & $0.69\pm0.03$ & $0.11\pm0.03$ & $2.23\pm0.03$ & $0.16\pm0.03$ \\
		27 (G) & 03   52 00  & -68   24 00  & 8.7 & 266.6 & 267 & $222.2\pm1.3$ & $20.4\pm1.0$ & $1.83\pm0.01$ & $0.13\pm0.01$ & $-0.58\pm0.01$ & $0.12\pm0.01$ \\
		28 (G) & 04   06 00  & -62   30 00  & 10.9 & 301.3 & 303 & $210.9\pm1.1$ & $19.1\pm0.9$ & $1.78\pm0.01$ & $0.16\pm0.01$ & $-0.49\pm0.01$ & $0.10\pm0.01$ \\
		29 (G) & 07   16 00  & -66   06 00  & 10.6 & 84.6 & 266 & $324.7\pm1.1$ & $17.6\pm0.8$ & $1.29\pm0.01$ & $0.10\pm0.01$ & $1.40\pm0.01$ & $0.12\pm0.01$ \\ \hline
	\end{tabular}
\end{adjustbox}
\end{table*}

MagES utilises the 2dF multi-object fibre positioner in combination with the dual-beam AAOmega spectrograph on the AAT. The 2dF positioner allows for the observation of \textasciitilde350 science targets per two-degree diameter field. We configure the blue arm of AAOmega with the 1500V grating, which has resolution R\textasciitilde3700, to provide coverage of the MgIb triplet, and the red arm with the 1700D grating, which has  R\textasciitilde10000, to provide coverage of the near-infrared CaII triplet. Reduction of the spectra is performed using the \oldstylenums{2}\textsc{dfdr} pipeline \citep{aaosoftwareteam2dfdrDataReduction2015}, and line-of-sight (LOS) velocities are derived using cross-correlation against template spectra. After applying quality cuts, stars with heliocentric velocity estimates are cross-matched against the Gaia EDR3 catalogue, and additional cuts based on Gaia parameters \textaltfont{ruwe}<1.4 \citep{fabriciusGaiaEarlyData2021} and $C^*$<$4\sigma_{C^*}$\footnote{$C^*$ and $\sigma_{C^*}$, describing the corrected flux excess and thus the consistency between the $G$, $G_{\text{BP}}$, and $G_{\text{RP}}$ photometric bands, are defined using Eqs. 6 and 18 of \citet{rielloGaiaEarlyData2021} respectively.} applied.

\subsection{Field kinematics}\label{sec:kindata}
The resulting sample of stars includes both true Magellanic stars and foreground contaminants. We use a statistical framework, described in detail in \citetalias{C20}, to probabilistically associate stars to either the Clouds, or to one of several possible Milky Way contaminant populations based on their kinematics. These association probabilities are used to weight the fitting of a multi-dimensional Gaussian distribution describing the aggregate Magellanic kinematic properties of each field: the LOS velocity ($V_{\text{LOS}}$) and proper motions ($\mu_\alpha$, $\mu_\delta$)\footnote{$\mu_\alpha$ refers to proper motion in the $\alpha$ direction with the usual $\cos(\delta)$ correction, i.e. \textaltfont{PMRA} from the Gaia EDR3 source catalogue.}, as well as the dispersion in each of these components ($\sigma_{\text{LOS}}, \sigma_\alpha, \sigma_\delta$). Fitting is performed using the Markov Chain Monte Carlo ensemble sampler \textsc{emcee} \citep{foreman-mackeyEmceeMCMCHammer2013} to sample the posterior distribution of the parameters describing the Gaussian model. We report the 68 per cent credible interval as the 1$\sigma$ uncertainty in each of the six fitted parameters. Table~\ref{tab:fields} provides the inferred kinematic properties for each field analysed in the present work, as well as the number of stars in the field with an individual probability $P_i$>$50$\% of being associated with the Clouds. 

Several fields, particularly those located in the southern outskirts of the LMC, could plausibly contain more than one Magellanic population -- that is, include stars associated with both the LMC and SMC. To test this idea, we recalculate the above fits assuming a Magellanic population described by two Gaussian components, and compute the Bayesian Information Criterion \citep[BIC:][]{schwarzEstimatingDimensionModel1978} for comparison to that for the fit with a single component. We find that in all cases there is insufficient evidence to prefer the two-Gaussian fit over the single-component fit, and so therefore retain the single-component fit for all fields. We note this does not preclude the existence of both LMC and SMC stars within a field, but simply that any such populations are either i) not kinematically distinct, or ii) strongly mismatched in size (i.e. one population dominates over the other). 

In many fields, the number of likely Magellanic stars is significantly lower than the total number of stars observed. In the case of “M” fields (see \citetalias{C20} for details of these classifications), this is primarily due to the relatively inefficient target selection used, as these fields were observed prior to the release of Gaia DR2. The associated lack of proper motion and parallax information available for the target selection process, in combination with the moderate Milky Way contamination within the CMD selection boxes used to isolate Magellanic red clump stars (see Fig.~2 of \citetalias{C20}), means a significant fraction of the targets observed in these fields are not genuinely Magellanic members. In contrast, MagES fields observed after the release of Gaia DR2 (classified as “G” fields) do incorporate these kinematic priors, and as a result generally suffer less from contamination by non-members: for regions of comparable underlying target density, the target selection efficiency for “G” fields is approximately double that for “M” fields. Fields 3, 9, and 24-26 also have comparatively low numbers of likely Magellanic stars, despite their classification as “G” fields. This is due to the inherently low density of Magellanic stars at these locations, as seen in Fig.~\ref{fig:map}. 

\subsubsection{Field 3}\label{sec:sf1}	 
Notably, we find that the Magellanic population in field 3 is likely associated with the SMC, rather than the LMC. Comparison of its proper motions to those of nearby fields reveals it is very similar to fields 2 and 4, located in the SMC outskirts (identified in Fig.~\ref{fig:map} but not explicitly discussed in this paper), and kinematically distinct from fields 6 and 9, each of which have proper motions significantly more likely to be LMC-associated. In addition, the mean metallicity of this field is [Fe/H]\textasciitilde$-1.4$: somewhat lower than each of the other analysed fields (discussed further below), and suggestive of a potential origin in the more metal-poor SMC. As such, while we report the observed properties of this field in Table~\ref{tab:fields}, we do not analyse it further in the context of the LMC in this paper. However, we do note that at a SMC galactocentric radius of $9.5^{\circ}$, this is one of the most distant detections of SMC debris to date. It is more remote than the stellar overdensity discussed in \citet{pieresStellarOverdensityAssociated2017a}, and may sample the trailing arm of the SMC \citep[cf.][]{belokurovCloudsStreamsBridges2017}. A detailed analysis of this field is deferred to a forthcoming paper on the SMC by the MagES collaboration (Cullinane et al. in prep). 

\subsection{Metallicities}\label{sec:mets}
MagES additionally reports [Fe/H] estimates for sufficiently bright ($G\gtrsim18$) red giant branch stars, derived from the equivalent width of the 8542\AA{} and 8662\AA{} CaII triplet lines. These bright RGB stars are not included in the target selection procedure for “M” fields, but are present in all “G” fields. \citetalias{C20} and \citet{dacostaCaIiTriplet2016} describe the equivalent width measurement and [Fe/H] conversion procedure in detail.

For fainter red clump stars (observed in all fields), the S/N is too low to accurately measure the equivalent width of the two lines in any individual stellar spectrum, particularly as the 8662\AA{} line is within a region of the spectrum relatively heavily contaminated by (stochastically over- or under-subtracted) night sky emission. We therefore stack spectra for likely ($P_i>50$\%) Magellanic RC stars -- first shifting these into the rest frame using their (geocentric) LOS velocities -- to create a single “representative” RC spectrum for the field. This increases the contrast of the two CaII lines relative to the residual night-sky emission, allowing for equivalent width measurements to be performed. We test varying the likelihood threshold $P_i$ used to define the stack, but find there are no systematic variations in resultant [Fe/H] estimates, indicating any potentially misclassified galactic contaminants in the stack do not significantly affect our results. As the stacked clump stars only occupy a small magnitude range (and thus, we assume, a small range in effective temperature and gravity), stacking spectra is not expected to substantially bias the derived equivalent widths, and the resulting [Fe/H] estimates are expected to tend towards the mean metallicity within a given field. All metallicity estimates have uncertainties of 0.2~dex (we refer the interested reader to \citetalias{C20} for details). 

\subsection{Red Clump properties}\label{sec:rcdat}
Converting the aggregate proper motion for a given field to a physical velocity, as well as placing constraints on the geometry of the outer LMC disk and surrounding features, requires a distance estimate for the Magellanic stars in each field. To obtain these, we use the red clump as a standardizable candle, calibrated to an empirical distance scale as described in the next Section. The properties of the red clump are measured using Gaia EDR3 photometry at the location of each field. In principle, these provide additional information about the thickness and composition of the disk. 

For each field, we select all\footnote{including those for which we do not have line-of-sight velocity measurements.} stars within a $1^\circ$ radius of the field centre, with parallax $\varpi$<$0.15$, proper motions within 3 times the dispersion of the field median motions reported in Table~\ref{tab:fields} (i.e. $\mu_\alpha\pm3\sigma_\alpha$, and $\mu_\delta\pm3\sigma_\delta$), and passing the quality cuts \textaltfont{ruwe}<1.4 and $C^*$<$4\sigma_{C^*}$. 

The Gaia G-band photometry is then corrected for the 6-parameter solution as described in \citet{rielloGaiaEarlyData2021}, and subsequently dereddened using the procedure described in \citet{gaiacollaborationGaiaEarlyData2021c}. This utilises the \citet{schlegelMapsDustInfrared1998} dust maps, corrected as described in \citet{schlaflyMeasuringReddeningSloan2011}, in conjunction with the mean extinction coefficients for the Gaia passbands described in \citet{casagrandeUseGaiaMagnitudes2018}. No correction is made for reddening internal to the Clouds as this is not expected to be significant in the low-density peripheral regions targeted by MagES \citep[cf.][]{choiSMASHingLMCTidally2018}. We do, however, restrict our selection to stars with corrected $E(B-V)$<$0.25$ in order to minimise the effect of any systematic uncertainties in the dereddening procedure on the resultant photometry. To isolate red clump stars, a CMD selection box of $0.8$<$(G_{BP}-G_{RP})_0$<$1.15$, and $18.0$<$G_0$<$19.4$ is implemented; the large range in $G_0$ is designed to accommodate any reasonable distance variations between fields. 

\begin{figure}
	\includegraphics[width=0.9\columnwidth]{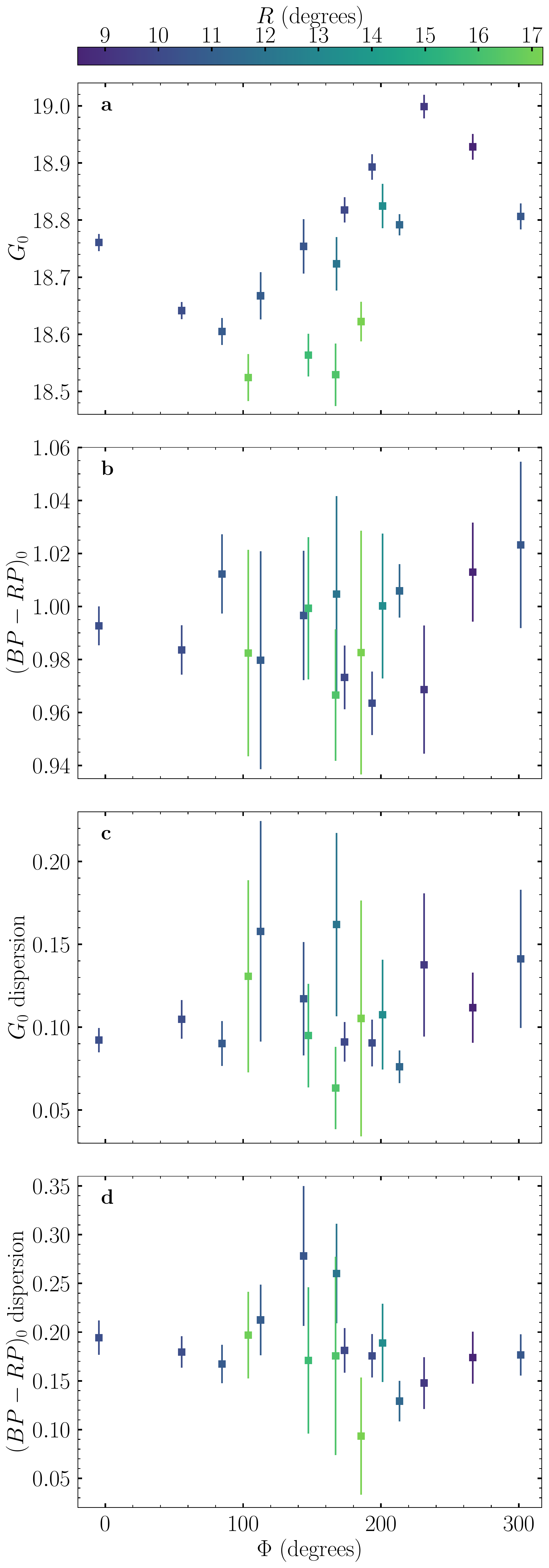}
	\caption{Fitted red clump parameters for MagES fields as a function of position angle. Panels show, in order from top to bottom: a) the peak $G_{0\text{c}}$ magnitude, b) the peak $(G_{BP}-G_{RP})_{0\text{c}}$ colour, c) the $G_0$ magnitude dispersion $\sigma_{G0}$, and d) the $(G_{BP}-G_{RP})_{0}$ colour dispersion $\sigma_{BPRP0}$. Points are colour-coded by their LMC galactocentric radius in degrees.}
	\label{fig:rcprops}
\end{figure}

To this final sample of stars we fit a mixture model describing their distribution on the CMD, within which the red clump takes the form of a two-dimensional Gaussian with five free parameters: a peak at ($G_{0\text{c}}$, $(G_{BP}-G_{RP})_{0\text{c}}$), dispersions $\sigma_{BPRP0}$ in the colour direction and $\sigma_{G0}$ in the magnitude direction, and a covariance $\sigma_{GBPRP0}$. In addition, the background density (accounting for both non-Magellanic populations, and non-RC Magellanic stars including RGB and potentially blue loop stars) is described by linearly varying terms in both colour and magnitude. The relative fraction of contaminants to true RC members is also a free parameter, for a total of 10 free model parameters. With on order of \textasciitilde100 stars per field, there is a sufficient number of data points to robustly determine each of these parameters. The Markov Chain Monte Carlo ensemble sampler \textsc{emcee} \citep{foreman-mackeyEmceeMCMCHammer2013} is used to sample the posterior probability distributions for the model parameters. We assuming uniform priors for each parameter, additionally requiring the location of the Gaussian peak to be within the CMD selection box. Fig.~\ref{fig:rcprops} presents the peak red clump colour ($(G_{BP}-G_{RP})_{0\text{c}}$) and magnitude ($G_{0\text{c}}$), and associated dispersions, for each field as a function of position angle. The approximately sinusoidal distribution of $G_{0\text{c}}$ as a function of position angle, with brighter magnitudes toward the northeast and fainter magnitudes toward the southwest, are broadly as expected for the inclined LMC disk. 

As noted above, all fields except for 25 and 26 have a sufficient number of stars to robustly determine the RC properties. For fields 25 and 26, the Magellanic stellar density is inherently very low, as evidenced by Table~\ref{tab:fields}: these fields have amongst the lowest numbers of likely Magellanic stars. For these two fields only, we therefore expand our RC selection criteria to include stars within 3$^\circ$ of the MagES field centre, which increases the number of stars sufficiently to fit the RC properties in these regions. We do note that as a consequence, it is possible that any distance gradients across these larger regions can potentially inflate $\sigma_{G0}$ beyond that associated with the line-of-sight depth of the LMC material within the field. Such an effect is expected to be less significant in the other analysed fields, which occupy smaller on-sky areas. 

\section{Anchoring the LMC red clump distance scale}\label{sec:fitdisk}
As a standardizable candle, the apparent magnitude of the red clump can be employed to derive distance estimates to each of our observed fields. However, even after dereddening, the apparent magnitude of the clump is not purely governed by its distance: population effects including the age and metallicity of red clump stars affect their intrinsic luminosity \citep[see][for a review]{girardiRedClumpStars2016}. Consequently, meaningful distance estimates to our fields require a “reference magnitude”: the apparent magnitude of a red clump, comprised of an identical stellar population to that of our fields in the LMC outskirts, at a known distance. 

However, the geometry of -- and therefore absolute distance to -- the outer LMC disk ($R$>8$^\circ$) is relatively unconstrained. Existing models of the LMC’s disk based on kinematic data -- such as those from \citet{gaiacollaborationGaiaEarlyData2021a,vandermarelThirdEpochMagellanicCloud2014,vasilievInternalDynamicsLarge2018b} and \citet{wanSkyMapperViewLarge2020} -- are dominated by data within LMC galactocentric radii of \textasciitilde8$^\circ$: smaller than that of even the innermost MagES fields. Given the perturbed appearance of the outer disk, assuming \textit{a priori} these same disk geometries extend to larger radii is not justified. However, in \citetalias{C20} it was found that the kinematics of MagES field 18 are consistent with predictions of a simple inclined disk model, indicating there are potentially unperturbed regions of the outer LMC disk suitable for calibration of distance estimates. Motivated by this, we seek to fit an inclined disk model to the LMC at galactocentric radii of \textasciitilde10$^\circ$, which we can use in conjunction with the measured field-aggregate red clump magnitudes to derive a “reference magnitude” as described above. 

Our model is that of a thin, rotating disk as described by the framework in \citet{vandermarelMagellanicCloudStructure2001} and \citet{vandermarelNewUnderstandingLarge2002} (henceforth referred to as vdM01 and vdM02 respectively). We fit four model parameters: the inclination ($i$) and position angle of the line of nodes (LON: $\Omega$)\footnote{the axis along which the plane of the disk intersects the plane of the sky.}, the rotational velocity $V_\theta$, and $G_0^{\text{RC}}$: the RC magnitude at the (fixed) centre distance of the LMC. We stress this magnitude is not the actual RC magnitude at the LMC centre, because the RC population at that location is very different to that in the outer disk; but is the magnitude of a population equivalent to that observed in the outer LMC disk, were such a population located at the LMC centre distance of 49.59~kpc \citep{pietrzynskiDistanceLargeMagellanic2019}. 

The model additionally requires as inputs the location ($\alpha_0,\delta_0,D_0$) and kinematics ($V_{\text{LOS},0},\mu_{\alpha,0},\mu_{\delta,0}$) of the LMC centre: we hold these fixed at the values derived by \citet{gaiacollaborationGaiaEarlyData2021a} for stars with LMC radii >$3^\circ$, as this is a kinematic centre derived using similar stellar tracers to those we analyse. At the large galactocentric radii analysed, we assume that $V_\theta$ is fixed and does not vary with radius (see e.g. \citetalias{C20}), the mean radial $\overline{V_r}$ and vertical $\overline{V_z}$ velocities of the disk are zero (as expected for an undisturbed disk), and that there is no precession or nutation of the disk \citep[i.e. both the inclination and line of nodes are fixed: see e.g.][]{vandermarelThirdEpochMagellanicCloud2014}.

\begin{figure}
	\includegraphics[width=\columnwidth]{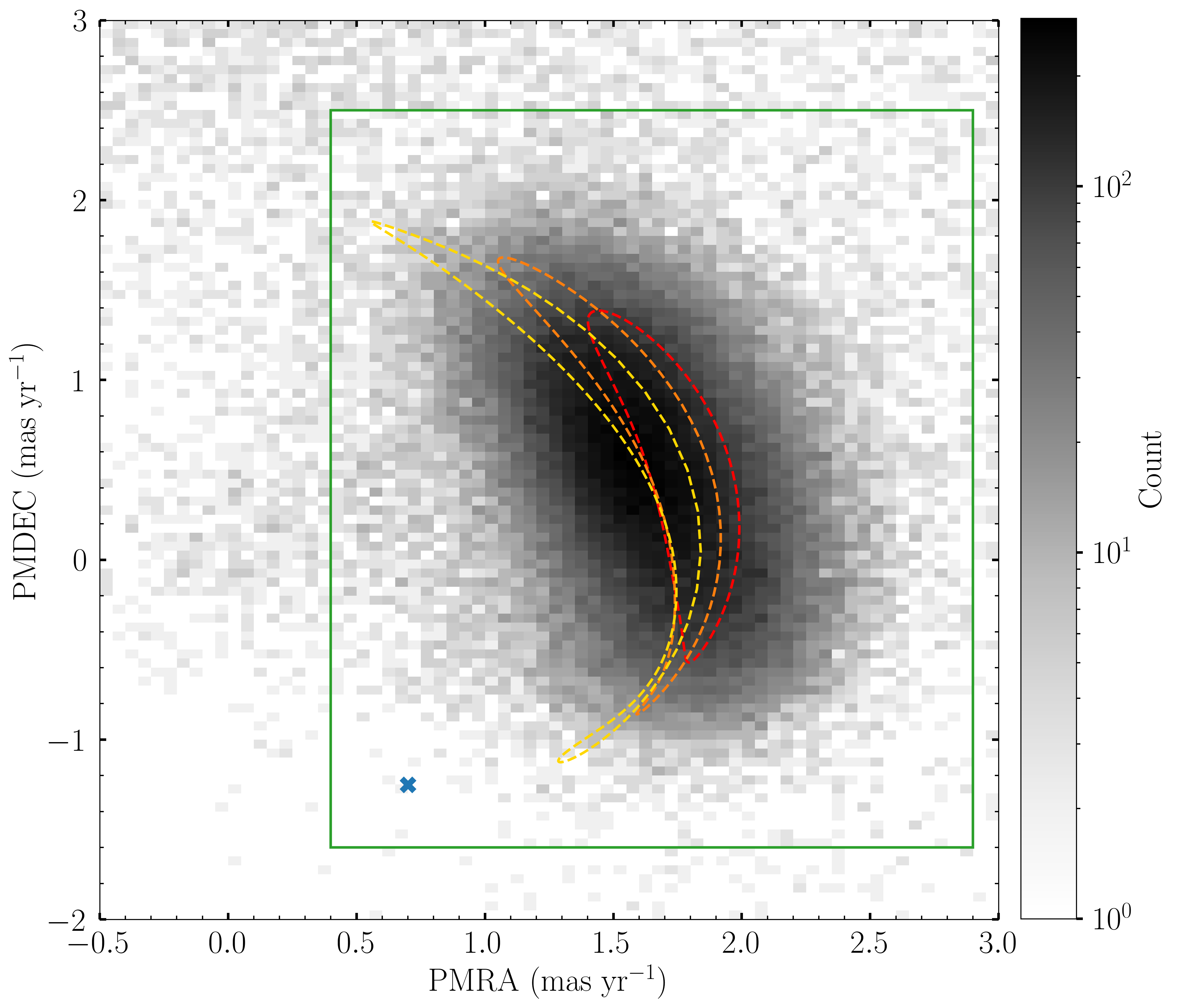}
	\caption{2D density plot of Gaia EDR3 proper motions for red clump stars (selected as described in \S\ref{sec:fitdisk}) at LMC radii between 7 and 12$^\circ$ of the LMC’s centre. Stars associated with the LMC predominantly form a distorted ellipse shape centred at \textasciitilde(1.5,0.6). A lower density peak at \textasciitilde(0,0) is associated with distant Milky Way halo stars with near-zero proper motions. The blue x-sign marks the mean proper motion of the SMC at \textasciitilde(0.7,$-$1.3), though as we only show stars at LMC radii <12$^\circ$, few SMC stars are present in this plot. Dashed lines show the predicted proper motion tracks for an LMC modelled as an inclined disk with geometry as described in \protect\citet{gaiacollaborationGaiaEarlyData2021a}, at galactocentric radii of 8$^\circ$ (red), 12$^\circ$ (orange), and 16$^\circ$ (gold). The solid green rectangle shows the proper motion selection for likely LMC disk stars used to fit the LMC disk geometry in \S\ref{sec:fitdisk}. This minimises contamination from the SMC and MW halo, whilst ensuring as complete as possible a sample of LMC stars: the shape of the distribution is such that elliptical selections would omit LMC stars at the large galactocentric radii of interest.}
	\label{fig:pmtrack}
\end{figure}

As MagES fields do not provide contiguous coverage of the LMC disk, to fit the model we instead select red clump stars (i.e. those within the CMD selection box described in Section~\ref{sec:rcdat}) from Gaia EDR3, passing the quality cuts \textaltfont{ruwe}<1.2 and $C^*$<$3\sigma_{C^*}$\footnote{These are stricter cuts than applied to general MagES photometry. This is because our aim here is to select the cleanest sample of Magellanic stars, whilst when studying individual fields, we aim to achieve a balance between both clean and complete photometry.}, to perform the fit. We additionally impose proper motion cuts of $0.4$<$\mu_\alpha$<$2.5$ and $-1.6$<$\mu_\delta$<$2.5$, and a parallax cut of $\varpi$<$0.15$ to isolate likely LMC stars, as seen in Fig.~\ref{fig:pmtrack}. We choose stars with LMC galactocentric radii of $9.5^{\circ}$<$R$<$10.5^{\circ}$ in order to have a consistent RC population with MagES disk fields (most of which have similar LMC galactocentric radii), and to minimise potential population effects such as radial metallicity gradients \citep[e.g.][]{carreraMetallicitiesAgeMetallicityRelationships2011,majewskiDiscoveryExtendedHalolike2008a} on the magnitude of the red clump. We later test the validity of the assumption that the RC population is consistent across MagES fields. 

As we require the RC magnitude to contribute to the fitting of model distances, we bin the data in position angle ranges of $2.5^{\circ}$, and take the median $G_0$ magnitude of each bin as the RC magnitude for use in fitting the model. We employ this simpler method, as compared to Gaussian fitting the RC in each bin as described in Section~\ref{sec:rcdat}, in order to simplify the overall fitting process. However, we do test the difference between the explicit RC fitting and the simple median $G_0$ magnitude, and find for well-populated bins\footnote{i.e. containing a large number of genuine Magellanic RC stars, as is the case for the LMC disk as opposed to nearby low-density substructures.}, there is no significant difference between the RC magnitude derived using the two methods. For consistency, we then use the median proper motions of each bin, and associated uncertainties, in the fitting process. 

To fit the model, we define a chi-squared quantity $\chi^2$ as in Eq.~\ref{eq:chisq}. Here, $\mu_{\alpha,\text{obs},j}$, $\mu_{\delta,\text{obs},j}$ and $m_{\text{obs},j}$ are the observed proper motions and $G_0$ magnitude for each bin $j$; $\mu_{\alpha,\text{err},j}$, $\mu_{\delta,\text{err},j}$ and $m_{\text{err},j}$ are the standard errors in each of those quantities. The quantities $\mu_{\alpha,\text{mod},j}$, $\mu_{\delta,\text{mod},j}$ and $m_{\text{mod},j}$ are the proper motions and $G_0$ magnitude predicted by the model for each bin, and are related to the four fitted model parameters through the framework of \citetalias{vandermarelMagellanicCloudStructure2001} and \citetalias{vandermarelNewUnderstandingLarge2002}. In particular: 

\begin{itemize}
	\item The LOS distance to a given bin, $D_{{\text{mod},j}}$, is given by Eq.~8 of \citetalias{vandermarelMagellanicCloudStructure2001}. The apparent magnitude of the bin is therefore given by $m_{\text{mod},j}=5\log_{10}\left(\frac{D_{{\text{mod},j}}}{D_0}\right) + G_0^{\text{RC}}$. 
	\item The 3D velocity vector of a given bin is given by Eqs.~11, 13, and 21 of \citetalias{vandermarelNewUnderstandingLarge2002}. Eqs.~7-8 of \citetalias{vandermarelNewUnderstandingLarge2002} are used to calculate the two proper motion components $\mu_{\alpha,\text{mod},j}$ and $\mu_{\delta,\text{mod},j}$ from this velocity, with the LOS velocity component not used in the fitting process. 
	\item Since we define a positive $V_\theta$ as describing clockwise rotation on the sky, in the same sense as observed in the LMC disk, we set $s=-1.0$ in Eq.~21 of \citetalias{vandermarelNewUnderstandingLarge2002}.
\end{itemize}

\begin{multline}\label{eq:chisq}
\chi^2 =  \sum_{j}\biggl[\left(\frac{\mu_{\alpha,\text{obs},j}-\mu_{\alpha,\text{mod},j}}{\mu_{\alpha,\text{err},j}}\right)^2 + \left(\frac{\mu_{\delta,\text{obs},j}-\mu_{\delta,\text{mod},j}}{\mu_{\delta,\text{err},j}}\right)^2 \\
{}+ \left(\frac{m_{\text{obs},j}-m_{\text{mod},j}}{m_{\text{err},j}}\right)^2\biggr] 
\end{multline}

The likelihood function for the model is thus defined as $\exp(-0.5\chi^2)$, and we sample the posterior probability distributions for the model parameters (maximising the log-likelihood) utilising the Markov Chain Monte Carlo ensemble sampler \textsc{emcee} \citep{foreman-mackeyEmceeMCMCHammer2013}, assuming uniform priors for each parameter. 

Initially, we utilise the full range of position angles around the LMC disk to fit the model. However, we find the resulting model parameters produce a poor fit (i.e. have significant residuals) across much of the disk (see Fig.~\ref{fig:resid} in Appendix~\ref{sec:appendixeq}). This indicates our simple model is not an accurate description of the entire outer disk. Consequently, and after some experimentation, we limit our fitting to position angles between $5^\circ$<$\Phi$<$90^\circ$. This region of the disk is not in the immediate vicinity of any known substructures or obvious overdensities -- which are indicative of potential perturbations -- and includes MagES field 18, which was previously found in \citetalias{C20} to be kinematically undisturbed. 

The resultant model parameters for this subset of position angles, presented in Table~\ref{tab:fitprops}, produce a significantly improved fit to the data within this position angle range, with much smaller residuals (see Fig.~\ref{fig:resid} in Appendix~\ref{sec:appendixeq}). The derived inclination and position angle of the line of nodes are similar to literature measurements of the LMC disk at smaller radii derived from both photometric and kinematic fits \citep[such as from][]{gaiacollaborationGaiaEarlyData2021a,vasilievInternalDynamicsLarge2018b,vandermarelThirdEpochMagellanicCloud2014,olsenWarpLargeMagellanic2002,vandermarelMagellanicCloudStructure2001}, though we do note our inclination is towards the higher end of the literature range, particularly compared to some purely photometric measurements \citep[e.g.][]{choiSMASHingLMCTidally2018,subramanianStructureLargeMagellanic2013,koerwerLargeMagellanicCloud2009}. The derived rotation velocity is also consistent with literature values (see also \citetalias{C20}), and using our model parameters on the MagES disk fields with position angles between $5^\circ$<$\Phi$<$90^\circ$ reveals these fields obey the expected disk-like kinematics (i.e. $\overline{V_r}\sim\overline{V_z}\sim0$: see Section~\ref{sec:3dkins}) even when LOS velocity information is included. This indicates our assumption that the north-eastern region of the LMC disk is relatively kinematically unperturbed is reasonable, and our method provides a viable reference red clump magnitude.

\begin{table}
	\caption{Fitted LMC disk model parameters, for position angles between $5^\circ$<$\Phi$<$90^\circ$.}
	\label{tab:fitprops}
	\begin{adjustbox}{max width=\columnwidth}
	\begin{tabular}{llll}
		\hline
		Inclination ($i$,$^\circ$) & Position angle of the LON ($\Omega$,$^\circ$) & $V_\theta$ (km~s$^{-1}$) & $G_0^{\text{RC}}$  \\ \hline
		$36.5\pm0.8$ & $145.0\pm2.5$ & $69.9\pm1.7$ & $18.90\pm0.01$  \\ \hline
	\end{tabular}	
	\end{adjustbox}
\end{table}

\section{Stellar populations in the LMC outskirts}\label{sec:pops}
A critical assumption underpinning our derivation of a “reference” red clump luminosity in Section~\ref{sec:fitdisk} is that the stellar populations across the outer LMC disk and associated MagES fields are largely homogeneous, such that distance variations are the dominant factor driving differences in the observed red clump magnitude field to field. Here we demonstrate the validity of this assumption by assessing the extent to which age and metallicity may vary in the LMC outskirts given our observations of the stellar populations in this region, and how such variations could affect the red clump luminosity.

\begin{figure}
	\includegraphics[width=\columnwidth]{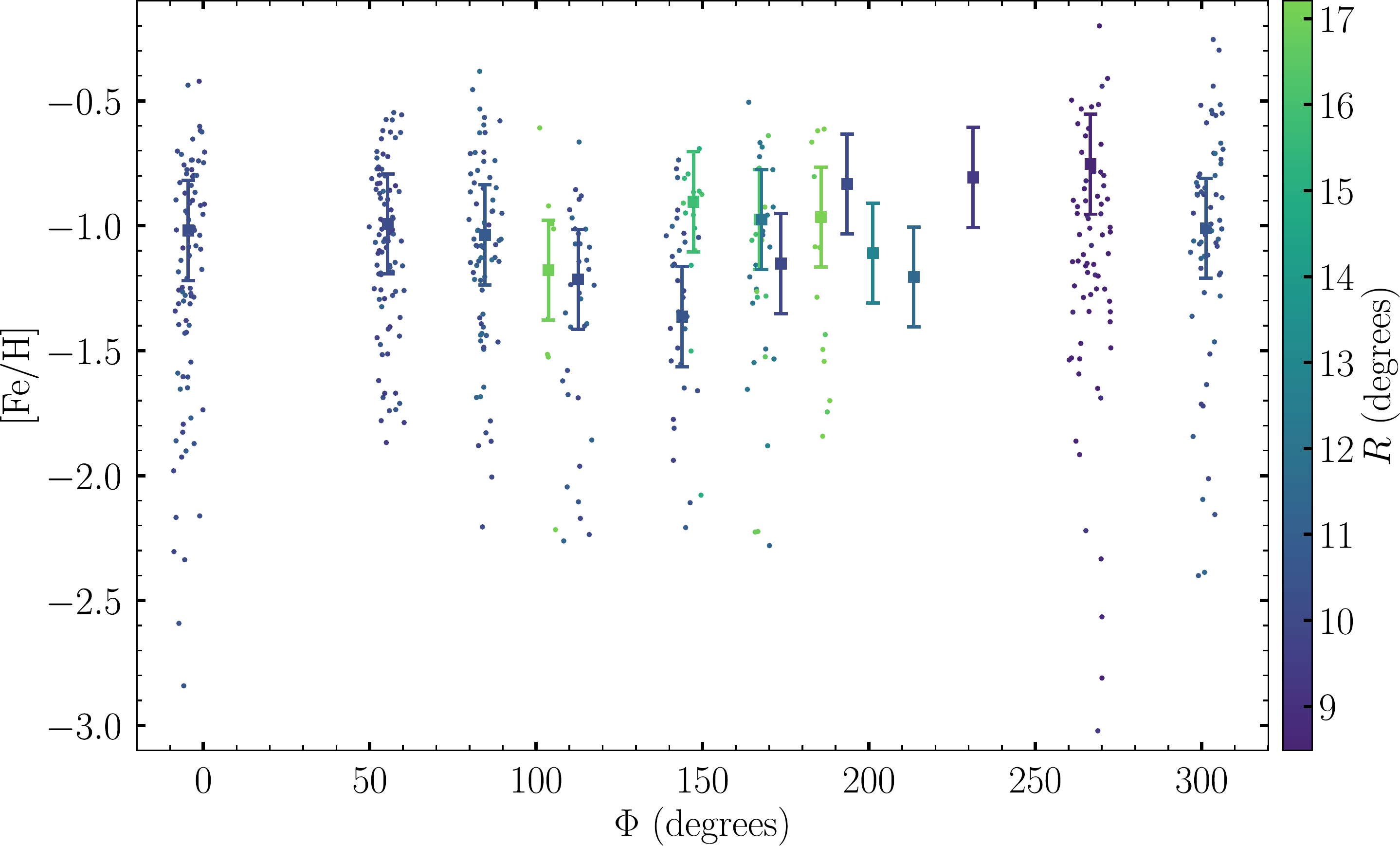}
	\caption{[Fe/H] estimates for MagES fields as a function of position angle. Fields are colour-coded by their galactocentric radius from the LMC. Square points represent results from stacked spectra, which approximate the mean metallicity of the field. Errorbars represent 0.2~dex abundance uncertainty. Dots represent [Fe/H] estimates for individual RGB stars within each field; associated uncertainties are omitted for clarity. “M” fields in the southern LMC disk do not include RGB stars, and therefore do not have associated [Fe/H] estimates for individual stars. }
	\label{fig:feh}
\end{figure}

We first assess the effect of metallicity, as in optical photometric bands (such as Gaia $G$) metal-poor RC stars are intrinsically brighter and bluer than those that are metal-rich \citep{girardiPopulationEffectsRed2001}. Fig.~\ref{fig:feh} presents [Fe/H] measurements for stars within MagES fields as a function of position angle and LMC galactocentric radius; square points represent results from stacked red clump spectra, which should approximate the mean metallicity within the field. We find these mean metallicities are, within uncertainty, generally consistent. The median stack metallicity across all fields is [Fe/H]$=-1.01$, with a standard deviation of 0.19. There are no systematic variations in the mean metallicity with galactocentric radius or position angle. We additionally find that, for fields where [Fe/H] estimates of RGB stars are available, the metallicity dispersion within each field is similar, with standard deviations of \textasciitilde0.5~dex. Consequently, we can infer that metallicity is not a dominant influence on the measured properties of the red clump in MagES fields. Our results are also broadly consistent with literature measurements in the outer LMC disk. Photometric metallicity estimates derived by \citet{gradyMagellanicMayhemMetallicities2021} using Gaia DR2 data find mean [Fe/H] values between $-1$ and $-1.5$ across the outskirts of the LMC disk, with dispersions averaging \textasciitilde0.5~dex (though we note their derived dispersions vary significantly from point-to-point). Sparse spectroscopic measurements by \citet{carreraMetallicitiesAgeMetallicityRelationships2011} and \citet{majewskiDiscoveryExtendedHalolike2008a}, predominantly in the northern outskirts of the disk, also find mean metallicities of approximately $-1$~dex. 

Stellar age also affects the properties of the red clump: young ($\lesssim2$ Gyr) RC stars are significantly bluer and brighter than older RC stars \citep{girardiPopulationEffectsRed2001}. However, DECam photometry in the LMC outskirts shows no evidence for significant populations of main sequence stars above the ancient \citep[\textasciitilde11~Gyr:][]{mackeySubstructuresTidalDistortions2018} turnoff across the regions studied in this paper, implying a similar lack of young RC stars. We can thus infer age is neither a dominant, nor systematic, influence on the measured properties of the red clump in MagES fields. 

Secondary constraints on population effects are also obtainable through analysis of the red clump colour, as changes in age and metallicity also drive changes in this parameter. Panel b of Fig.~\ref{fig:rcprops} shows the fitted $(G_{BP}-G_{RP})_{0}$ colour of the red clump in each of the MagES fields. We find these are, within uncertainty, consistent at $(G_{BP}-G_{RP})_{0}\sim1.0$, with no systematic variation in colour as a function of either radius or position angle. The dispersion in colour, presented in panel d of Fig.~\ref{fig:rcprops}, is similarly consistent, suggesting a very similar mix of populations across these locations.

The consistent metallicity and RC colour across the MagES fields we are considering, and lack of younger stellar populations in associated photometry at these locations, is indicative that population effects minimally impact the red clump in the LMC outskirts. To quantify this, we use PARSEC isochrones \citep{bressanParsecStellarTracks2012}\footnote{Accessed as version 3.4 of the web form: \url{http://stev.oapd.inaf.it/cmd}} to test the predicted colour and magnitude variations in Gaia EDR3 passbands associated with [Fe/H] variations at the level of the field-to-field dispersion in mean metallicity of \textasciitilde0.2~dex. The isochrones, for an 11~Gyr population assuming the default parameters for IMF and mass loss, predict that such changes result in $(G_{BP}-G_{RP})_{0}$ colour variations on the order of \textasciitilde0.06, consistent with the maximum colour difference across our observations. The implied change in intrinsic red clump luminosity is \textasciitilde0.04~mag, comparable to the \textasciitilde0.03~mag uncertainties on the RC magnitude of each field resulting from the fitting process described in Section~\ref{sec:rcdat}. We therefore conclude that distance is indeed the dominant factor impacting the observed RC magnitude. This validates our derivation of a “reference magnitude” for use in the calculation of distance estimates to our fields, and additionally allows us to use the magnitude dispersion of the red clump to draw conclusions about the relative line-of-sight thickness of the disk and nearby substructures. 

\section{Kinematics in the frame of the LMC disk}\label{sec:3dkins} 
Having ascertained that the north-eastern LMC outskirts retain a relatively intact disk structure in \S\ref{sec:fitdisk}, and having validated our assumption of largely homogeneous stellar populations across all the fields we consider in \S\ref{sec:pops}, we proceed to calculate the kinematics and distances of the fields relative to the frame of the LMC disk. We begin by deriving absolute distances to each field as in Eq.~\ref{eq:dist}, where $D_{\text{field}}$ is the distance in kpc to each field, $D_0$ is the distance of the reference red clump magnitude \citep[$49.59\pm0.63$~kpc:][]{pietrzynskiDistanceLargeMagellanic2019}, $G_0^{\text{RC}}$ is the reference RC magnitude from Table~\ref{tab:fitprops}, and $G_{0,\text{field}}$ is the peak RC magnitude for each field as determined in Section \ref{sec:rcdat}. We account for uncertainties in each of $D_0$, $G_{0,\text{field}}$ and $G_0^{\text{RC}}$, in derivation of associated uncertainties in $D_{\text{field}}$.

\begin{equation}\label{eq:dist}
	D_{\text{field}} = D_0\times10^{\left[\left(G_{0,\text{field}}-G_0^{\text{RC}}\right)/5.0\right]}
\end{equation}

We then utilise the framework presented in \citet{vandermarelNewUnderstandingLarge2002}, in conjunction with the derived distances to each field, to transform the observed field kinematics presented in Table~\ref{tab:fields} into velocities in a cylindrical coordinate system aligned with the LMC disk, with its origin at the LMC centre of mass. In this transformation, we utilise the LMC centre and associated systemic motions reported in \citet{gaiacollaborationGaiaEarlyData2021a}, and the disk inclination and LON angles derived in Section~\ref{sec:fitdisk} for the outer LMC disk. We subsequently obtain $V_\theta$, the azimuthal streaming or rotation velocity; $V_r$, the in-plane radial velocity; $V_z$, the vertical velocity perpendicular to the disk plane; and dispersions ($\sigma_\theta, \sigma_r, \sigma_z$) in each of these components. We additionally obtain the out-of-plane distance $z$, describing how far “in front of” or “behind” the expected disk plane stars in each field are located. For fields in the north-eastern disk, these distances are approximately zero by construction, since this is the region used to fit the disk plane (albeit without including line-of-sight velocity information). However, fields outside of this region are not required to -- and in fact do not -- lie within the expected disk plane\footnote{i.e. as data outside the region $5^\circ$<$\Phi$<$90^\circ$ is not used to fit the disk plane, the behaviour of the LMC disk outside this region (such as potential warps, etc.) is not necessarily well-described by the expected plane.}. Table~\ref{tab:vtrz} reports the derived disk velocity components and associated dispersions, as well as the out-of-plane distance, for each MagES field analysed.

\begin{figure*}
	\includegraphics[width=\textwidth]{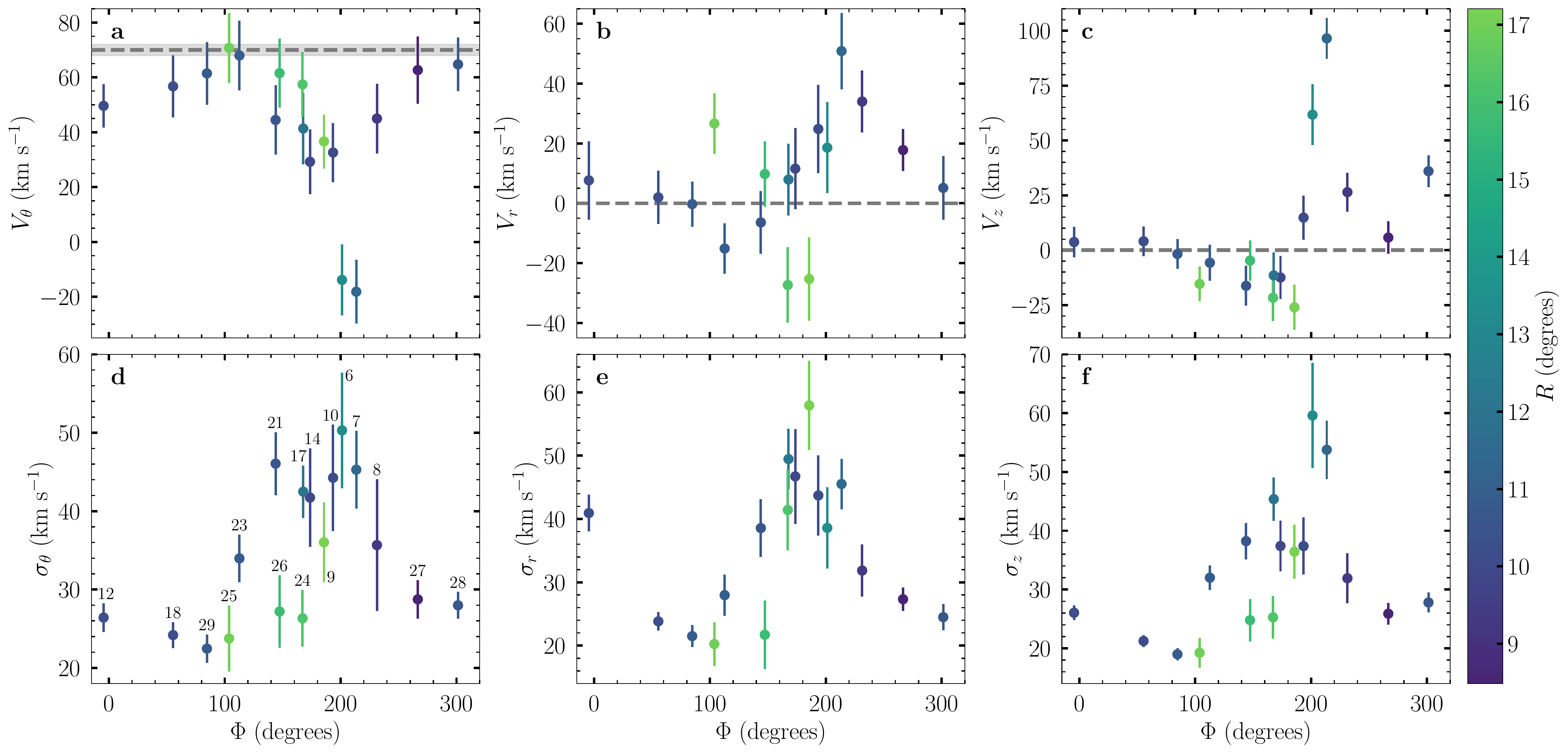}
	\caption{Azimuthal ($V_\theta$, left), radial ($V_r$, centre), and vertical ($V_z$, right) velocities (top row), and their associated dispersions (bottom row), for MagES fields as a function of position angle. Fields are coloured by their LMC galactocentric radius in degrees, and labelled with their associated field number in the lower left panel for reference. Dashed grey lines represent the expected kinematics for stars in an equilibrium disk.}
	\label{fig:diskvels}
\end{figure*}

\begin{figure}
	\includegraphics[width=\columnwidth]{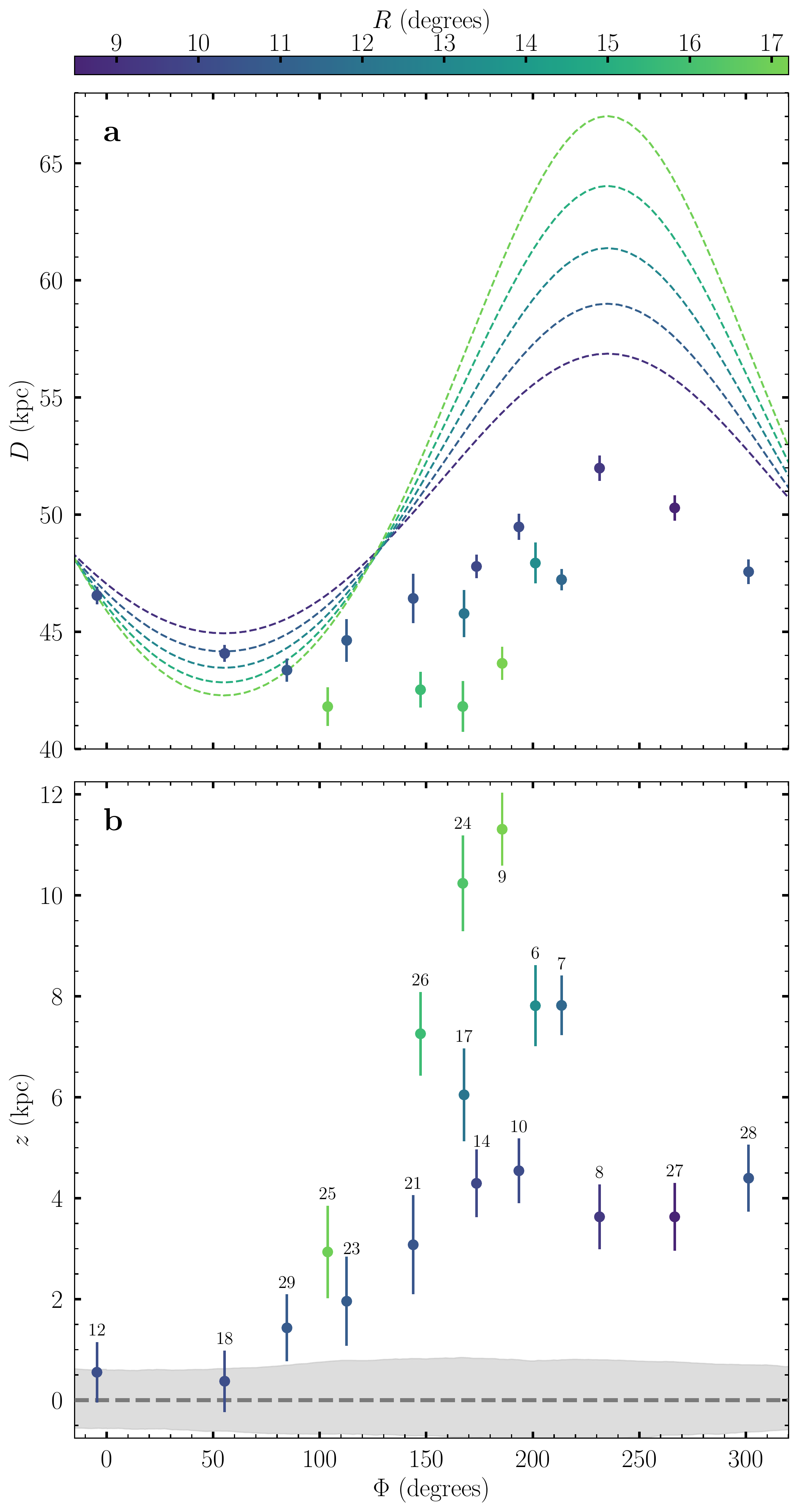}
	\caption{Absolute distances $D_{\text{field}}$ (top), and out-of-plane distances $z$ (bottom) relative to the LMC disk plane, for MagES fields as a function of position angle. Fields are coloured by their LMC galactocentric radius in degrees. Coloured dashed lines represent the distance predictions based on the inclined disk fit to the north-eastern LMC in \S\ref{sec:fitdisk} at the corresponding radii. The dashed grey line in the bottom panel represents the $z=0$ expectation for stars within the disk plane, with the shaded grey region indicating the distance uncertainty associated with uncertainties in the fitted disk inclination and line of nodes. Points in the lower panel are labelled with their associated field numbers for reference.}
	\label{fig:zdash}
\end{figure}

In this coordinate system, a positive $V_\theta$ refers to clockwise motion, from north towards west, following the sense of rotation in the LMC; a positive $V_r$ refers to motion in the disk plane radially outward from the LMC centre; and a positive $V_z$ refers to motion perpendicular to the disk plane in a direction predominantly toward the viewer (i.e. “in front of” the LMC disk). For stars obeying equilibrium disk kinematics, the net $V_r$ and $V_z$ within a field are expected to be zero, $V_\theta$ is expected to be \textasciitilde70~km~s$^{-1}$ as derived in Section~\ref{sec:fitdisk}\footnote{We note that the values derived for the north-eastern MagES disk fields are slightly lower than this, though still consistent within uncertainty, due to the inclusion of LOS velocities in this calculation, which are not available for the fitting process in \S\ref{sec:fitdisk}.}, and the out-of-plane distance is expected to be zero. Fig.~\ref{fig:diskvels} plots the resultant disk velocities and their dispersions for all analysed MagES fields, as a function of their position angle around the disk. Points are colour-coded by their radial distance from the LMC centre. Fig.~\ref{fig:zdash} plots both the absolute distances to the fields, as well as their out-of-plane distances relative to the LMC disk with geometry assumed as above. 

\begin{table*}
	\caption{Disk velocities and associated dispersions for MagES fields analysed in this paper. We also report the out-of-plane distance ($z$) for each field.}
	\label{tab:vtrz}
	\begin{tabular}{lrrrrrrr}
		\hline
		\multicolumn{1}{c}{Field} & \multicolumn{1}{c}{$V_{\theta}$ (km~s$^{-1}$)} &
		\multicolumn{1}{c}{$\sigma_\theta$ (km~s$^{-1}$)} & 
		\multicolumn{1}{c}{$V_{r}$ (km~s$^{-1}$)} & 
		\multicolumn{1}{c}{$\sigma_r$ (km~s$^{-1}$)} & 
		\multicolumn{1}{c}{$V_z$ (km~s$^{-1}$)} &
		\multicolumn{1}{c}{$\sigma_z$ (km~s$^{-1}$)} & \multicolumn{1}{c}{$z$ (km)}  \\ \hline
			6 & $-13.9\pm12.9$ & $50.3\pm7.4$ & $18.6\pm15.3$ & $38.6\pm6.4$ & $61.7\pm13.9$ & $59.6\pm8.9$ & $7.8\pm0.8$ \\
			7 & $-18.1\pm11.6$ & $45.3\pm5.0$ & $50.9\pm12.8$ & $45.5\pm4.0$ & $96.5\pm9.4$ & $53.8\pm5.0$ & $7.8\pm0.6$ \\
			8 & $45.0\pm12.7$ & $35.7\pm8.4$ & $34.0\pm10.4$ & $31.8\pm4.1$ & $26.4\pm8.8$ & $31.9\pm4.3$ & $3.6\pm0.6$ \\
			9 & $36.6\pm9.8$ & $36.0\pm5.1$ & $-25.3\pm13.9$ & $57.9\pm7.1$ & $-26.0\pm10.3$ & $36.4\pm4.6$ & $11.3\pm0.7$ \\
			10 & $32.6\pm10.8$ & $44.3\pm6.8$ & $24.8\pm14.7$ & $43.7\pm6.3$ & $14.8\pm10.1$ & $37.4\pm4.9$ & $4.5\pm0.6$ \\
			14 & $29.2\pm11.8$ & $41.7\pm6.3$ & $11.6\pm13.6$ & $46.7\pm7.5$ & $-12.5\pm9.8$ & $37.4\pm4.3$ & $4.3\pm0.7$ \\
			17 & $41.3\pm13.0$ & $42.5\pm3.4$ & $7.9\pm12.0$ & $49.5\pm4.8$ & $-11.5\pm10.6$ & $45.4\pm3.7$ & $6.0\pm0.9$ \\
			21 & $44.5\pm12.6$ & $46.1\pm4.0$ & $-6.4\pm10.5$ & $38.6\pm4.6$ & $-16.2\pm9.1$ & $38.2\pm3.1$ & $3.1\pm1.0$ \\
			12 & $49.6\pm7.9$ & $26.4\pm1.8$ & $7.7\pm13.1$ & $40.9\pm2.9$ & $3.7\pm6.9$ & $26.0\pm1.2$ & $0.6\pm0.6$ \\
			18 & $56.7\pm11.3$ & $24.2\pm1.7$ & $2.0\pm8.9$ & $23.8\pm1.5$ & $4.0\pm6.8$ & $21.2\pm1.0$ & $0.4\pm0.6$ \\
			23 & $68.0\pm12.7$ & $34.0\pm3.0$ & $-15.1\pm8.5$ & $28.0\pm3.3$ & $-5.7\pm8.2$ & $32.0\pm2.1$ & $2.0\pm0.9$ \\
			24 & $57.4\pm11.8$ & $26.3\pm3.6$ & $-27.3\pm12.6$ & $41.4\pm6.4$ & $-21.7\pm10.6$ & $25.3\pm3.7$ & $10.2\pm1.0$ \\
			25 & $70.7\pm12.8$ & $23.8\pm4.2$ & $26.6\pm10.1$ & $20.2\pm3.5$ & $-15.4\pm7.9$ & $19.2\pm2.5$ & $2.9\pm0.9$ \\
			26 & $61.5\pm12.6$ & $27.2\pm4.6$ & $9.8\pm10.9$ & $21.7\pm5.4$ & $-4.8\pm9.1$ & $24.8\pm3.6$ & $7.3\pm0.8$ \\
			27 & $62.6\pm12.3$ & $28.7\pm2.5$ & $17.8\pm7.1$ & $27.3\pm1.8$ & $5.7\pm7.4$ & $25.9\pm1.9$ & $3.6\pm0.7$ \\
			28 & $64.7\pm9.8$ & $28.0\pm1.7$ & $5.1\pm10.7$ & $24.5\pm2.1$ & $36.0\pm7.2$ & $27.8\pm1.7$ & $4.4\pm0.7$ \\
			29 & $61.4\pm11.5$ & $22.5\pm1.8$ & $-0.3\pm7.6$ & $21.5\pm1.7$ & $-1.7\pm6.7$ & $19.0\pm1.1$ & $1.4\pm0.7$ \\ \hline
	\end{tabular}
\end{table*}

\subsection{Common properties and literature comparison}\label{sec:common}
In the following subsections, we discuss the derived distances and kinematics of each of the individual regions of the LMC outskirts with distinct kinematic and structural properties. However, possible interpretations of these in the context of interactions between the Clouds are deferred to Section~\ref{sec:analysis}. We first examine common properties across all studied fields. 

With consistent red clump colours, consistent metallicity means and dispersions, and kinematics similar to those predicted by LMC disk models (see Fig.~\ref{fig:pmtrack}), we can infer that stars within all MagES fields analysed here\footnote{note this excludes field 3 as discussed in \S\ref{sec:sf1}.} are comprised predominantly of LMC disk populations. While we cannot preclude the existence of any SMC stars within our fields, particularly those in the southern LMC outskirts, we can conclude that the majority of stars in each field do not originate with the SMC, as such material would i) be more metal-poor, and ii) have kinematics more consistent with the known SMC proper motion field (as in the case of field 3 as discussed in Section~\ref{sec:sf1}). We can further infer that the majority of stars in our fields are not from an LMC stellar halo, which would be expected to not only be more metal-poor, but also have a lower rotation velocity and significantly larger velocity dispersions in all components than the LMC outer disk. While we cannot completely preclude the presence of LMC halo stars within our fields, the majority of stars are likely disk populations, even at the very large galactocentric radii studied. 

\citet{youssoufiStellarSubstructuresPeriphery2021} provide distance estimates to several of the observed structures in the LMC disk outskirts using infrared photometry from the VISTA Hemisphere Survey (VHS) and the VMC survey \citep{cioniVMCSurveyStrategy2011}. In that study, RC magnitude estimates are obtained by fitting the dereddened K-band luminosity function of red clump and RGB stars within a defined CMD selection box using a Gaussian distribution (which describes the clump) plus a quadratic polynomial term (which describes the RGB). The magnitudes are converted to distance estimates using the absolute K-band magnitude of solar-neighbourhood RC stars, with a correction factor from \citet{salarisPopulationEffectsRed2002} applied to account for the difference in stellar population properties (particularly metallicity) between the two samples. 

We thus compare our distance estimates to those provided by \citet{youssoufiStellarSubstructuresPeriphery2021}. Specifically, our fields 17 and 14 correspond to their “Southern Substructure 1”; fields 6 and 7 correspond to their “Southern Substructure 2”; fields 9 and 24 correspond to their “Southern Substructure 3”; and field 25 corresponds to their “Eastern Substructure 2”. We find the distances estimates from our analysis are systematically smaller than those presented in \citet{youssoufiStellarSubstructuresPeriphery2021}, being on average $7.5\pm1.1$~kpc closer, though the relative distances of the different substructures are consistent within uncertainty. At the distance of the LMC, this global offset corresponds to a magnitude difference of \textasciitilde0.35~mag, larger than our typical photometric uncertainties by a factor of \textasciitilde10. 

The most likely source of this global distance offset is a calibration difference. In particular, we note the \citet{salarisPopulationEffectsRed2002} magnitude correction $\delta M_{K}^{\text{RC}}=-0.07$ for the Clouds, used by \citet{youssoufiStellarSubstructuresPeriphery2021}, is derived for a younger and more metal-rich population \citep[{[Fe/H]}\textasciitilde$-0.4$: see Table 4 of][]{girardiPopulationEffectsRed2001}, which is more representative of the inner regions of the Clouds. However, as discussed in Section~\ref{sec:pops}, the outskirts of the Clouds are significantly more metal-poor, at [Fe/H]\textasciitilde$-1$. Since \citet{salarisPopulationEffectsRed2002} indicate more metal-poor RC stars (with ages>1.5~Gyr) are fainter than more metal-rich stars, this indicates the applied correction factor $\delta M_{K}^{\text{RC}}$ is likely underestimated. Fig.~6 of \citet{salarisPopulationEffectsRed2002} indicates for an old (\textasciitilde11~Gyr) population at this metallicity, a more appropriate correction factor is approximately $-$0.4: similar to the global magnitude offset between our estimates and those of \citet{youssoufiStellarSubstructuresPeriphery2021}. Indeed, applying this alternate correction factor to the \citet{youssoufiStellarSubstructuresPeriphery2021} photometry results in distances consistent within uncertainty to those we derive. 

However, we do also note also that the \citet{youssoufiStellarSubstructuresPeriphery2021} distance estimates are calculated for much larger on-sky areas than individual MagES fields, with even their smallest comparable region (Southern Substructure 2) encompassing an on-sky area of \textasciitilde25 square degrees. This is \textasciitilde8x larger than the 3 square degrees covered by MagES fields. As a result, the \citet{youssoufiStellarSubstructuresPeriphery2021} structures have very large associated distance dispersions, on the order of 8~kpc: similar to the offset between the two sets of distance estimates. It is therefore plausible, though we suspect very unlikely, that our fields simply sample the closer end of the distance distribution across each substructure region. 

\begin{figure}
	\centering
	\raisebox{-0.5\height}{\includegraphics[width=0.8\columnwidth]{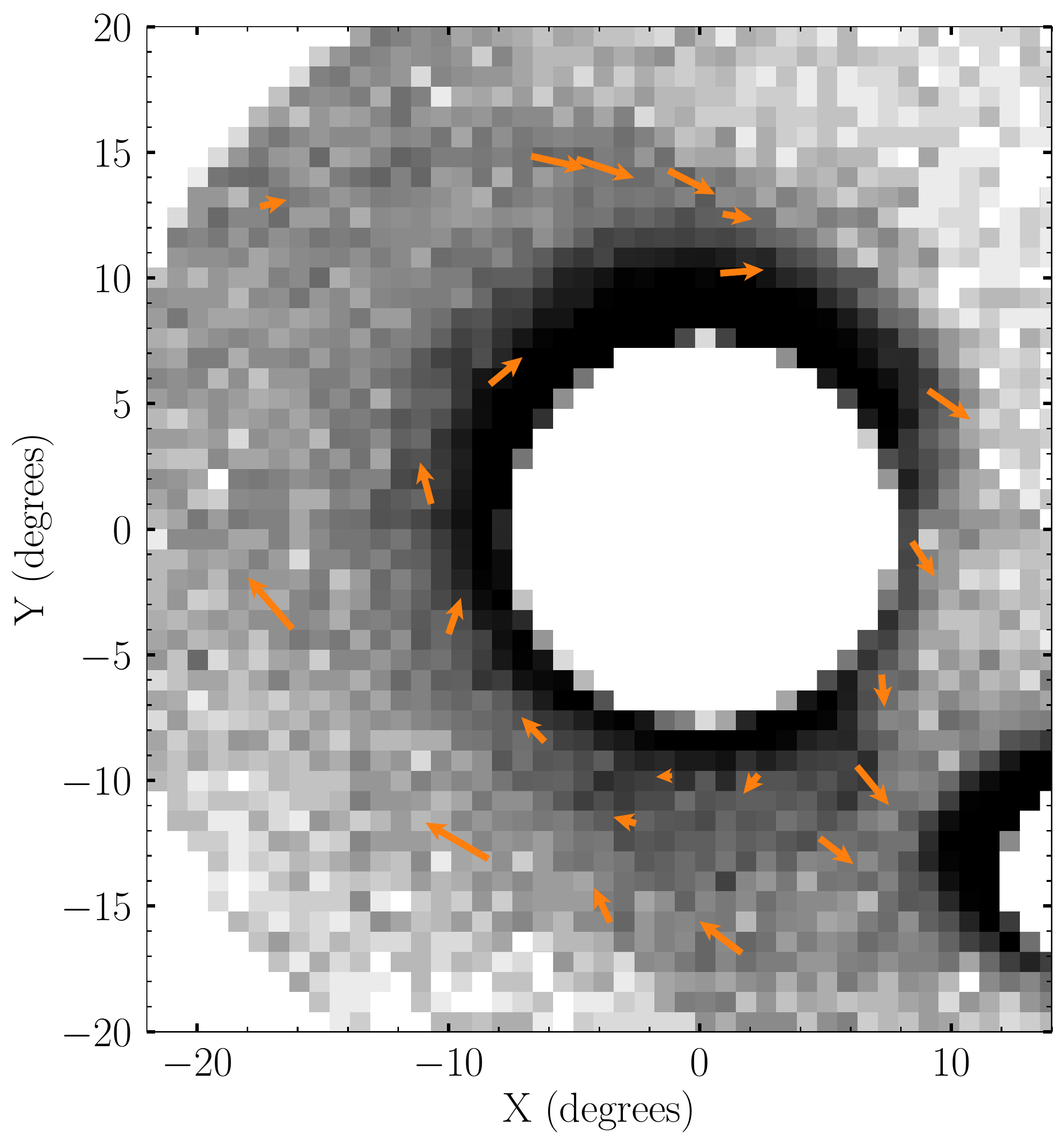}}
	\raisebox{-0.47\height}{\includegraphics[width=0.185\columnwidth]{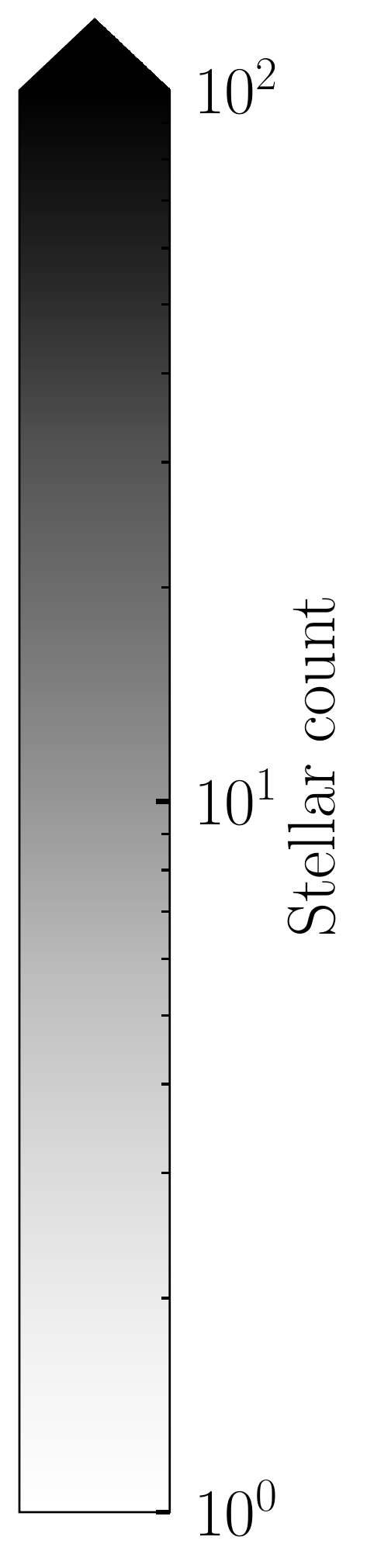}}
	\caption{Proper motion vector field (orange arrows) for MagES fields in the outskirts of the LMC. For completeness, in addition to the MagES fields which are the focus of this work, we also include measurements for fields along the LMC's northern arm taken from \protect\citetalias{C21}, though this feature is not discussed further in this paper. The background image shows the stellar density of likely Magellanic stars with LMC galactocentric radii >7$^\circ$ and SMC galactocentric radii >$4^\circ$ selected from Gaia EDR3 using similar criteria to \citet{gaiacollaborationGaiaEarlyData2021a}. Uncertainties in the length and direction of the vectors due to proper motion uncertainties (not shown) are on the order of \textasciitilde20\% and 5\% respectively.}
	\label{fig:vmap}
\end{figure}

We additionally briefly compare our kinematic results to the proper motion map of the Magellanic outskirts presented in Fig.~17 of \citet{gaiacollaborationGaiaEarlyData2021a}. To facilitate this comparison, we produce a similar velocity vector map to that figure in our Fig.~\ref{fig:vmap}, which presents the aggregate proper motions of each MagES field discussed in this paper. These are overlaid on a stellar density map of likely Magellanic red clump stars selected from Gaia EDR3 using similar criteria to those of the \textit{“LMCout”} sample in \citet{gaiacollaborationGaiaEarlyData2021a}\footnote{As a point of interest, we additionally present the proper motion vector field associated with the full sample in Appendix \ref{sec:appendixpm}.}. To obtain this vector field, we first subtract the LMC's bulk motion from each of the measured aggregate proper motions in Table~\ref{tab:fields}, accounting for projection effects as in \cite{vandermarelNewUnderstandingLarge2002}. This correction requires knowledge of the LOS distance to each field, as calculated using Eq.~\ref{eq:dist}, and is thus slightly different to that utilised by \citet{gaiacollaborationGaiaEarlyData2021a}. The remaining proper motion component is then projected into the $x_{\text{LMC}},y_{\text{LMC}}$ coordinate system described using Eqs.~1-2 of \citet{gaiacollaborationGaiaEarlyData2021a}.

Broadly, Fig.~\ref{fig:vmap} agrees with the kinematics displayed in Fig.~\ref{fig:diskvels}: each field displays azimuthal motion consistent with the LMC's sense of (clockwise) rotation, except for fields 6 and 7 (discussed further in Section~\ref{sec:claw}). This is also qualitatively consistent with Fig.~17 of \citet{gaiacollaborationGaiaEarlyData2021a}. Differences in the apparent length of individual vectors in Fig.~\ref{fig:vmap} are due primarily to differences in the LOS distances to each field. For example, as seen in the top panel of Fig.~\ref{fig:zdash}, fields along the southern arm-like feature are \textasciitilde7~kpc closer than fields in the southern LMC disk, and have correspondingly larger projected proper motions.
	
We also find the velocity field presented in our map appears somewhat less strongly radial in the LMC's southern and eastern outskirts, and somewhat more strongly radial along the LMC's western outskirts, than that of \citet{gaiacollaborationGaiaEarlyData2021a} -- particularly at the largest radii. However, these differences are attributable to the different prescriptions used to correct for the LMC's bulk velocity in the two maps.  

We do note a fundamental limitation of these maps is that they are simply projections of proper motion measurements, and thus cannot account for the contribution of the LOS velocity component to the full 3D kinematics of each field as in Fig.~\ref{fig:diskvels}. Since, at the large galactocentric radii of MagES fields, 
$V_{\text{LOS}}$ does not directly correspond to $V_z$ (and instead contributes to all three disk velocity components), the velocity vectors in Fig.~\ref{fig:vmap} do not perfectly align with the directions expected from Fig.~\ref{fig:diskvels}. We therefore caution over-interpretation of Fig.~\ref{fig:vmap}, and for the remainder of this paper discuss only the full 3D velocity field relative to the plane of the LMC disk.

\subsection{The relatively undisturbed north-eastern disk}\label{sec:northeast} 
As discussed in Section~\ref{sec:rcdat}, Gaia data reveals the north-eastern LMC disk is well-described by an inclined disk with an azimuthal (rotation) velocity of \textasciitilde70~km~s$^{-1}$. This is supported by the kinematics and structural properties of fields 12, 18 and 29, which span the north-eastern disk (position angles $-5^\circ$<$\Phi$<90$^\circ$). We remind the reader that the radial distances and position angles of all fields are given in Table~\ref{tab:fields}.

These fields have vertical and radial velocities consistent with zero within uncertainty, as expected in an equilibrium disk, and have velocity dispersions in each component of 20-25~km~s$^{-1}$, consistent with literature measurements of similar tracer populations at smaller galactocentric radii \citep{gaiacollaborationGaiaEarlyData2021a,vasilievInternalDynamicsLarge2018b}. While field 12 ($\Phi$\textasciitilde$-5^\circ$) is mildly deviant from these values, with an azimuthal velocity \textasciitilde1.5$\sigma$ lower than the canonical value and an elevated radial velocity dispersion of \textasciitilde40~km~s$^{-1}$, this field is located directly adjacent to the northern arm feature discussed in \citetalias{C21}. It is therefore unsurprising that minor deviation from equilibrium disk kinematics is observed, likely resulting from the same interactions that formed the nearby substructure. In particular, the elevated in-plane radial velocity dispersion is a known indicator of perturbation \citep{wanSkyMapperViewLarge2020}.

In terms of disk structure, fields 18 and 12 have out-of-plane distances consistent with zero within uncertainty, and the out-of-plane distance for field 29 is only just in excess of a 1$\sigma$ deviation from zero. We additionally find $\sigma_{G0}$, the red clump magnitude dispersion in each field (which is an indication of the line-of-sight thickness of the disk) is consistent across these fields, as expected given the consistent vertical velocity dispersions between these fields. 

\subsection{The “straight-edged” western disk}\label{sec:west}
Fields 8, 27, and 28 are each located in the western outskirts of the LMC disk (230$^\circ$<$\Phi$<330$^\circ$), which appears sharply truncated compared to the north-eastern disk in maps such as Fig.~\ref{fig:map}. Each of these fields are located \textasciitilde4~kpc in front of the predicted LMC disk plane, and it is thus not surprising that fields 8 and 28 have \textasciitilde30~km~s$^{-1}$ vertical velocities in this direction. The in-plane radial velocity is strongest in field 8, at \textasciitilde35~km~s$^{-1}$ and decreases moving north along the disk edge, reaching \textasciitilde5~km~s$^{-1}$ (consistent with zero within uncertainty) in field 28. We also find the velocity dispersion in each component decreases moving north from field 8 to field 28; however even in field 28 these velocity dispersions are uniformly \textasciitilde10~km~s$^{-1}$ larger than those in the undisturbed north-eastern disk (i.e., fields 18 and 29). Correspondingly, the red clump magnitude dispersion within each of these fields appears mildly elevated compared to the undisturbed disk, though we note the large uncertainties in $\sigma_{G0}$ mean this difference is not statistically significant. 

\subsection{The disturbed southern disk}\label{sec:south} 
The southern outskirts of the LMC disk, including fields 21, 14 and 10 (100$^\circ$<$\Phi$<220$^\circ$ and 10$^\circ$<$R$<11$^\circ$), are significantly disturbed from equilibrium disk kinematics. Velocity dispersions in these fields, typically 40-50~km~s$^{-1}$ in each component, are approximately double those in the undisturbed disk, and the azimuthal velocities of the fields are also \textasciitilde2$\sigma$ lower than that in the undisturbed disk. However, we do find that despite the large observed velocity dispersions, these fields do not display elevated red clump magnitude dispersions ($\sigma_{G0}$) compared to the undisturbed disk. This is at least partly attributable to projection effects, as $\sigma_{G0}$ traces only the LOS dispersion of the field. While in the northern outskirts of the LMC, the LOS direction closely maps the $V_z$ direction (such that an elevated vertical velocity dispersion is expected to be observed as an increased LOS velocity dispersion, and thus increased $\sigma_{G0}$), in the southern outskirts of the LMC the line-of-sight projects to a combination of all three (vertical, azimuthal, and radial) disk frame coordinates. Consequently, an increased velocity dispersion in any one of these individual directions will not necessarily introduce a significant increase in $\sigma_{G0}$. However, as we observe elevated velocity dispersions in all three components in these fields, it is perhaps surprising we do not see some increase in $\sigma_{G0}$ in these fields. 

Field 23, located in the south-eastern disk ($\Phi$\textasciitilde110$^\circ$), marks a transition from the relatively ordered kinematics of the north-eastern disk to the disturbed kinematics of the southern disk. The azimuthal and vertical velocities in this field are consistent within uncertainty with those expected for an equilibrium disk, unlike the other southern disk fields. However, its velocity dispersions -- each \textasciitilde30-35~km~s$^{-1}$ -- are elevated relative to the undisturbed disk, albeit still \textasciitilde1$\sigma$ less than the other southern disk fields. We note that while this field does have a large $\sigma_{G0}$ dispersion, uncertainty on this value is sufficiently large that it is still potentially consistent with the $\sigma_{G0}$ values measured in the nearby disk fields. 

All of the southern disk fields display clear kinematic trends in their radial and vertical velocities as a function of position angle. The in-plane radial velocity of the fields increases from approximately $-$20~km~s$^{-1}$ in field 23 to \textasciitilde25~km~s$^{-1}$ in field 10, with the adjacent field 8 in the south-western disk continuing the observed trend. In addition, the out-of-plane vertical velocity increases from approximately $-$15~km~s$^{-1}$ in field 21 to \textasciitilde15~km~s$^{-1}$ in field 10, with the adjacent field 8 continuing this trend. An increasing vertical velocity is potentially compatible with the observed increase in the out-of-plane distance from \textasciitilde2~kpc to \textasciitilde4.5~kpc between fields 21 and 8. Interestingly, we note that \citet{olsenWarpLargeMagellanic2002} -- who fit an LMC disk plane consistent with our measurements -- also find stars in the south-western LMC disk have positive out-of-plane distances of up to \textasciitilde2.5~kpc, though at much smaller galactocentric radii than our fields. 

At face value, our finding of positive out-of-plane distances for fields in the southern LMC outskirts appears inconsistent with results from \citet{choiSMASHingLMCTidally2018}, who find a warp towards the SMC (implying negative out-of-plane distances) based on photometry of the LMC red clump in this region. However, much of this apparent discrepancy is due to the higher inclination of the LMC disk in our analysis ($i$\textasciitilde$36.5^\circ$, compared to $i$\textasciitilde$25.86^\circ$ derived by \citealt{choiSMASHingLMCTidally2018}), relative to which the out-of-plane distance is calculated. More recent kinematic analysis from \citet{choiRecentLMCSMCCollision2022} find an LMC disk inclination of \textasciitilde35$^\circ$ -- consistent with our measurement -- for galactocentric radii $\gtrsim8^\circ$.
	
\subsection{The claw-like southern substructures}\label{sec:claw} 
The discovery of two “claw-like” substructures in the southern outskirts of the LMC is first described in \citet{mackeySubstructuresTidalDistortions2018}, with the more prominent “Substructure 2” located further west than “Substructure 1”. MagES fields are located on both substructures, with field 17 ($\Phi$\textasciitilde170$^\circ$) located on Substructure 1, and fields 6 and 7 ($\Phi$\textasciitilde205$^\circ$) located on Substructure 2 (see Fig.~\ref{fig:map}). We find the two features are kinematically distinct, and thus discuss each substructure individually. 

We compare the kinematics of field 17, located on Substructure 1, with those of field 14, located immediately adjacent in the southern LMC disk. The radial and vertical velocities in field 17 are very similar to those in field 14, and while the azimuthal velocity is \textasciitilde10~km~s$^{-1}$ higher in field 17, this is still consistent within 1$\sigma$ with that measured in 14. The azimuthal and radial velocity dispersions are also consistent between the fields. The vertical velocity dispersion is \textasciitilde8~km~s$^{-1}$ (\textasciitilde2$\sigma$) larger in field 17, consistent with a very mildly elevated $\sigma_{G0}$ in this field, though the large uncertainty on this value is such that this is not a statistically significant difference. The only substantial difference between the two fields is their out-of-plane distances: field 17 is \textasciitilde6~kpc in front of the undisturbed disk plane, a distance \textasciitilde2~kpc greater than field 14. We therefore suggest field 17 is likely perturbed by the same forces as the nearby southern disk. 

In contrast, fields 6 and 7, located on Substructure 2, have distinctively different kinematics to the neighbouring fields 10 and 8. Both fields 6 and 7 have enormous vertical velocities: \textasciitilde60~km~s$^{-1}$ in field 6, and \textasciitilde100~km~s$^{-1}$ in field 7. These are respectively 5 and 8$\sigma$ higher than that predicted for an equilibrium disk. This potentially explains the very large out-of-plane distances of these fields -- both are located \textasciitilde8~kpc in front of the predicted disk plane, and \textasciitilde4~kpc further than the nearby fields 10 and 8. The vertical velocity dispersion in the substructure is also 5$\sigma$ larger (at \textasciitilde55-60~km~s$^{-1}$) than the undisturbed disk, and 2$\sigma$ larger than the nearby disturbed disk (\textasciitilde35-40~km~s$^{-1}$). We note that despite the strongly elevated vertical velocity dispersion, $\sigma_{G0}$ in the substructure is not elevated relative to the nearby disk fields. This is because at this point in the disk, due to projection effects, the vertical velocity dispersion is contributed to most significantly by the dispersion in the PMRA direction (which as per Table~\ref{tab:fields}, is elevated in these fields relative to the nearby disk fields), and not the LOS direction. The azimuthal and radial velocity dispersions in these fields are not significantly different from those in the nearby southern disk (\textasciitilde45~km~s$^{-1}$), where we similarly do not see an elevated $\sigma_{G0}$, despite these velocity dispersions being elevated compared to the undisturbed disk. In sum, this suggests these fields are predominantly perturbed by a force in the $z$ direction, with possibilities for this discussed in greater detail in Section~\ref{sec:analysis}.  

Unexpectedly, as is evident in panel \textit{a} of Fig.~\ref{fig:diskvels}, fields 6 and 7 are also counter-rotating relative to the rest of the disk, with negative azimuthal velocities. While \citet{olsenPopulationAccretedSmall2011} also find a population of counter-rotating stars in the LMC, attributed to SMC debris infalling to the LMC potential, it is unlikely this is also the case for this substructure. As discussed above, the RC colour and mean [Fe/H] for the field are consistent with that of the LMC, while the \citet{olsenPopulationAccretedSmall2011} SMC population was found to be significantly more metal-poor than their LMC stars. Further, the \citet{olsenPopulationAccretedSmall2011} SMC population was found to be either counter-rotating in a plane closely aligned to that of the LMC disk, or co-rotating in a plane strongly inclined relative to the LMC disk. Both options are inconsistent with the counter-rotation observed at a very large out-of-plane distance in our fields. We therefore maintain the substructure is comprised of LMC disk material, significantly perturbed from equilibrium kinematics. 

\subsection{The extended southern substructure}\label{sec:southarm} 
The presence of a long, thin arm-like feature in the far southern outskirts of the LMC was first reported in \citet{belokurovCloudsArms2019}, and hypothesised as a potential counterpart to the northern arm. Four MagES fields are located along the locus of the feature: fields 9, 24, 25 and 26 (each at $R$\textasciitilde16.5). However, unlike the northern arm (discussed in \citetalias{C21}), these fields do not follow consistent kinematic or structural trends along the length of the feature. We thus present the properties of each field individually below, while discussion of implications for the perturbations producing the different features in each field are deferred to Section~\ref{sec:analysis}.

Field 25, located \textasciitilde3~kpc in front of the predicted LMC disk plane in the extreme eastern outskirts of the LMC ($\Phi$\textasciitilde100$^\circ$), is in the vicinity of “Eastern Substructure 2” in \citet{youssoufiStellarSubstructuresPeriphery2021} \citep[noted also in][]{gaiacollaborationGaiaEarlyData2021a}. This field has an in-plane radial velocity of \textasciitilde25~km~s$^{-1}$, and a vertical velocity of approximately $-$15~km~s$^{-1}$: each of which are \textasciitilde$2\sigma$ from the expected kinematics in an equilibrium disk. In contrast, we find the velocity dispersions and red clump magnitude dispersion $\sigma_{G0}$ for the field, in addition to its azimuthal velocity, are consistent within uncertainty with those in the undisturbed disk. 

Field 26 is located in the extreme south-eastern outskirts of the LMC ($\Phi$\textasciitilde150$^\circ$). Kinematically, it is consistent within uncertainty with the predicted equilibrium disk kinematics, with only mildly elevated velocity dispersions (\textasciitilde5~km~s$^{-1}$: <1$\sigma$ significant) relative to the undisturbed disk, and a consistent red clump magnitude dispersion also. However, this field is located \textasciitilde7~kpc in front of the predicted LMC disk plane, indicative of a perturbation, which is difficult to reconcile with the effectively undisturbed kinematics in the field.

In contrast to the relatively ordered kinematics of fields 25 and 26, fields 9 and 24 (165$^\circ$<$\Phi$<185$^\circ$) are more significantly disturbed. Both fields have in-plane radial velocities of approximately $-$25~km~s$^{-1}$, indicating inward motion towards the LMC centre. This is the opposite direction to velocities measured in the southern disk and claw-like substructures, but is similar to the negative in-plane radial velocities observed along the northern arm-like feature discussed in \citetalias{C21}. The fields additionally have vertical velocities of approximately $-$25~km~s$^{-1}$, despite being located $\geq$10~kpc in front of the predicted LMC disk plane. This is unlike the nearby claw-like Substructure 2, which is similarly located significantly in front of the disk plane, but has correspondingly large and positive vertical velocities. This suggests that different perturbations are likely responsible for the two structures. 

The velocity dispersions in these two fields are elevated relative to those in the undisturbed disk, with magnitudes similar to those in the southern disk fields, though field 24 is only significantly elevated in its radial velocity dispersion (\textasciitilde40~km~s$^{-1}$). Field 9 has elevated dispersions in each component, but is also most significantly elevated in its radial velocity dispersion (\textasciitilde55~km~s$^{-1}$). Correspondingly, the red clump magnitude dispersion in field 9 is larger than that in field 24, though the large uncertainties in $\sigma_{G0}$ due to the comparatively low number of stars in these fields mean these are not statistically significant differences. These elevated velocity dispersions are also unlike the northern arm, which is kinematically cold \citepalias[with velocity dispersions typically on the order of $\lesssim$25~km~s$^{-1}$: see][]{C21}. 

\section{Discussion}\label{sec:analysis} 
\subsection{Dynamical models}\label{sec:models} 

In order to assist in our interpretation of the disturbed kinematics in the LMC outskirts, we compare our results to a suite of simple dynamical models of the Magellanic system first presented in \citetalias{C21}. A brief summary of the key model parameters is presented below. We note that these models are designed only as a first exploration of the large and complex parameter space that describes the allowable orbits of the Clouds, and there are consequently associated model limitations. Most significant of these is likely the lack of self-gravity incorporated into the models, which can affect both the orbits of the Clouds, and the response of stars within the Clouds to close interactions (see \citetalias{C21} for a detailed discussion of these caveats). As such, here we only perform qualitative, rather than quantitative, comparisons to our observations. We note that even these qualitative comparisons may be biased by the simplistic nature of the models, though as discussed in \citetalias{C21}, these limitations most strongly affect very close interactions between the Clouds, which are relatively infrequent in our models (see discussion below). More detailed models are required to validate and expand upon our analysis, particularly regarding origins of substructures.

We model the LMC as a collection of test particles within a 2-component potential: an exponential disk with mass $2\times10^9$~M$_\odot$, scale radius 1.5~kpc, and scale height 0.4~kpc; and a Hernquist \citep{hernquistAnalyticalModelSpherical1990} dark matter halo of mass $1.5\times10^{11}$~M$_\odot$ \citep{erkalTotalMassLarge2019} and scale radius 20~kpc, such that the circular velocity is \textasciitilde90~km~s$^{-1}$ at 10~kpc \citepalias{C20}. The test particle distribution within the disk is initialised using \textsc{AGAMA} \citep{vasilievAGAMAActionbasedGalaxy2019} to account for the velocity dispersion of the LMC disk. The SMC is modelled as a Hernquist profile with mass $2.5\times10^9$~M$_\odot$ and scale radius 0.043~kpc, such that the SMC has a circular velocity of 60~km~s$^{-1}$ at 2.9~kpc \citep[motivated by the results of][]{stanimirovicNewLookKinematics2004}\footnote{As the entire SMC mass is enclosed within this radius in our models, this results in much smaller scale radii than in e.g. \citet{beslaRoleDwarfGalaxy2012a}, who model an initially more massive SMC which experiences mass loss through repeated interactions with the LMC.}. We do not initialise the SMC potential with tracer particles. The Milky Way is modelled as a 3-component system with a bulge, disk, and dark matter halo similar to the \textaltfont{MWPotential2014} from \cite{bovyGalpyPythonLibrary2015}. As in \citet{erkalTotalMassLarge2019}, we treat each of the three systems in the model (i.e. the MW, LMC, and SMC) as a particle sourcing a potential, allowing us to account for the motion of the Milky Way in response to the LMC \citep[as in][]{gomezItMovesDangers2015}. We account for the dynamical friction of the Milky Way on the LMC using the results of \citet{jethwaMagellanicOriginDwarfs2016a}, but do not explicitly account for the effects of dynamical friction between the LMC and SMC. 

The LMC and SMC are initialised at their present day locations, then rewound for 1~Gyr in the presence of each other and the Milky Way. At this time, the LMC disk is initialised with \textasciitilde$2.5\times10^6$ tracer particles, with a geometry matching that from \citet{choiSMASHingLMCTidally2018}. Consequently, when calculating velocities in the frame of the disk for the models, we use the that same geometry to define the disk plane. The system is then evolved to the present. 

We note the \citet{choiSMASHingLMCTidally2018} inclination ($i=25.86^\circ\pm1.4^\circ$) used in the models is somewhat lower than that we derive in \S\ref{sec:rcdat} ($i=36.5^\circ\pm0.8^\circ$), though the position angle of the LON ($\Omega=149.23^\circ\pm8.35^\circ$) is consistent within uncertainty with that we derive ($\Omega=145.0^\circ\pm2.5^\circ$). This is because these model suites were originally designed for study of the LMC's northern arm in \citetalias{C21}, which found the \citet{choiSMASHingLMCTidally2018} geometry is broadly consistent with the measured distance gradient along the arm (though note that paper also found a higher-inclination disk of \textasciitilde34$^\circ$ is also compatible with the measured distance gradient). We discuss the implications of the model disk geometry in further detail below (\S\ref{sec:inc_discuss}), but find our conclusions are generally not sensitive to small changes in the assumed inclination.

In \citetalias{C21}, we run multiple model suites in order to probe the range of allowed masses for the LMC, SMC, and MW; but in this paper, we discuss only two in detail: 

\begin{itemize}
\item Our fiducial or “base-case” model suite, which utilises our best estimates for the masses of the LMC, SMC, and Milky Way; and
\item A “no-SMC” suite, which omits the SMC, but is otherwise identical to the base-case suite. This allows for separation of the effects of the SMC from those of the MW. 
\end{itemize}
Findings from our model suites with increased MW and SMC masses are qualitatively similar to that of our base-case suite.

Within each model suite, we run multiple individual realisations, sampling from literature uncertainties on the current-day distances and systemic velocities of both the LMC and SMC, to explore the allowable parameter space. This results in differing orbits, and thus interaction histories, for the Clouds. As discussed in \citetalias{C21}, in all model realisations, the SMC has had a recent close pericentric passage around the LMC \textasciitilde150~Myr ago \citep[in agreement with][]{zivickProperMotionField2018b}, with a total pericentric distance $r_{\text{peri}}=8.0^{+2.4}_{-2.0}$~kpc. This distance is consistent with work by \citet{choiRecentLMCSMCCollision2022}, who find an impact parameter $\lesssim$10~kpc is necessary for the recent SMC pericentre to produce the observed level of disk heating in the inner LMC. However, in our models these pericentres occur significantly below the plane of the LMC disk\footnote{i.e. behind the disk plane relative to us.}, with $z_{\text{peri}}=-6.8^{+2.5}_{-2.6}$~kpc: the SMC is only approximately now crossing the LMC disk plane\footnote{Note that given the recent timing of the pericentric passage, this is broadly consistent with the \citet{choiRecentLMCSMCCollision2022} assertion that the pericentric passage is coincident with the SMC crossing the LMC disk plane.}. The projected galactocentric radius of the pericentric passage is \textasciitilde4~kpc, in a direction toward the southwest of the LMC. 

At earlier times, the orbit of the SMC varies significantly depending on how the systemic motions of both Clouds are sampled. Approximately 51\% of our base-case realisations have a second SMC crossing of the LMC disk $400^{+85}_{-70}$~Myr ago, which can occur across a broad range ($28.8^{+11.4}_{-9.2}$~kpc) of in-plane radial distances. The remaining \textasciitilde49\% of orbits remain behind the LMC’s disk plane for the 1~Gyr over which our models are run, and include some orbit realisations where the SMC is on its first infall to the LMC potential.

A handful of model realisations (\textasciitilde9\%) additionally have a third disk crossing $900^{+60}_{-160}$~Myr ago, though a significantly larger fraction would experience this crossing if our models were rewound for a greater length of time than 1~Gyr. The particulars of this crossing are much less robustly constrained than the \textasciitilde400~Myr crossing, with crossing distances of $53.8^{+13.1}_{-46.3}$~kpc permitted due to the increasing uncertainty in the SMC’s orbit at earlier times. We additionally find a small fraction (\textasciitilde4\%) of our models show a second SMC pericentric passage at around this time, again noting this fraction would significantly increase were our models rewound further than 1~Gyr. These pericentres have similar distances ($r_{\text{peri}}=6.2^{+3.8}_{-2.3}$~kpc) to the most recent pericentric passage, but occur at smaller out-of-plane distances ($z_{\text{peri}}=2.2_{-1.1}^{+2.5}$~kpc) due to the similarly-timed disk crossing in these realisations. 

\begin{figure*}
	\centering \includegraphics[width=\textwidth]{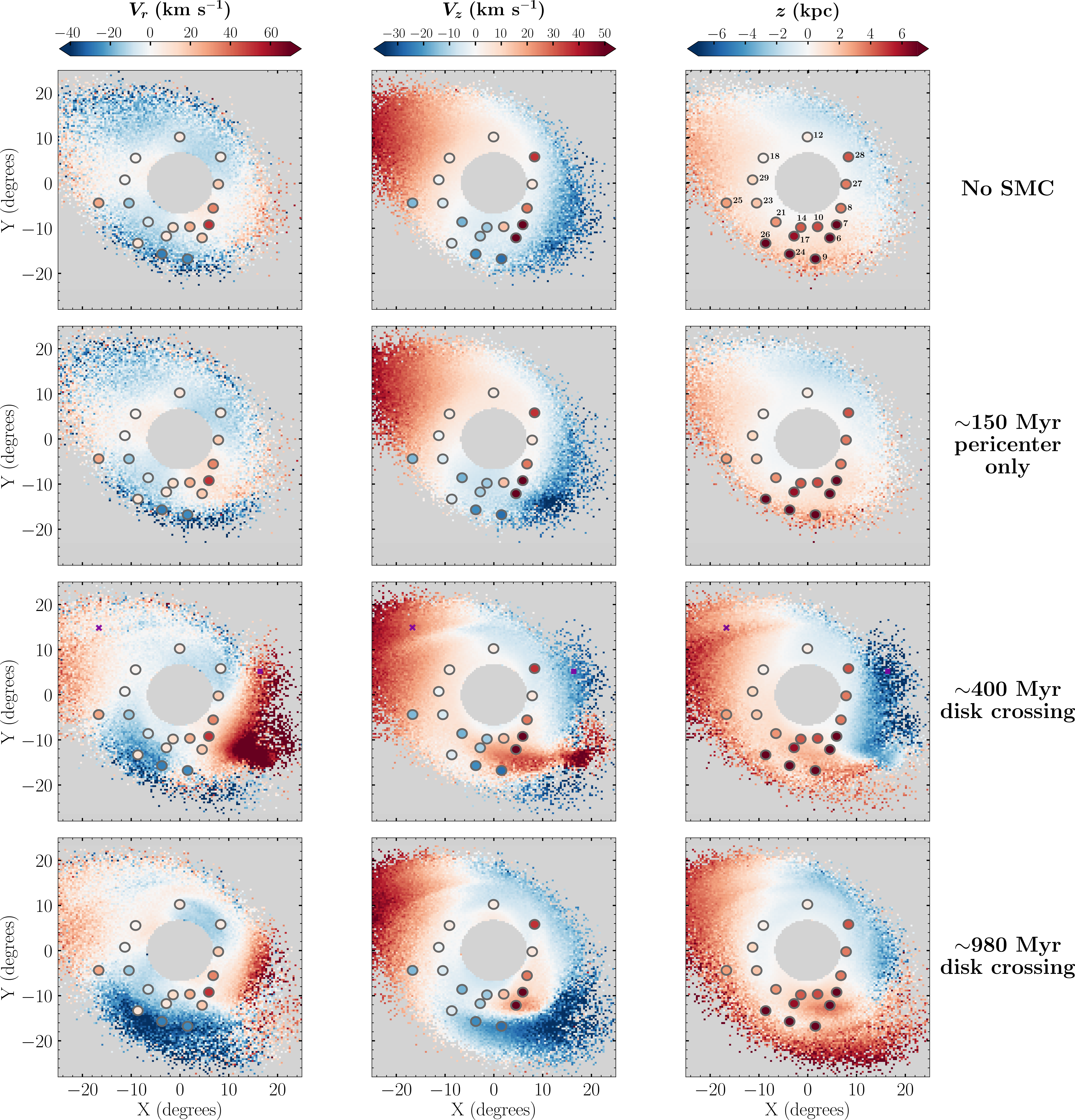}
	\caption{
			Kinematic predictions for four individual model realisations with differing LMC-SMC orbital histories, but similar LMC-MW orbital histories. Rows show model realisations which, in order from top to bottom: i) exclude the SMC entirely, ii) experience only an SMC pericentre \textasciitilde150~Myr ago, iii) experience an SMC crossing of the LMC disk plane \textasciitilde400~Myr ago and the SMC pericentre \textasciitilde150~Myr ago, and iv) experience SMC crossings of the LMC disk plane at both \textasciitilde980 and \textasciitilde400~Myr ago, in addition to the SMC’s pericentre \textasciitilde150~Myr ago. Columns show, in order from left to right: i) the in-plane radial velocity ($V_r$), ii) the out-of-plane vertical velocity ($V_z$), and iii) the out-of-plane distance ($z$). Grey circles represent the locations of the MagES fields studied in this analysis, with the fill colour within each circle indicating the observed kinematics of that field. The centre 7$^\circ$ of each model is masked to emphasise the kinematics of the outskirts. The top right panel indicates MagES field numbers, for reference. In the third row, the location of the \textasciitilde400~Myr crossing, at the crossing time, is marked by purple x-signs, with the approximate present-day location -- computed by rotating the location of the original crossing within the LMC’s disk plane, assuming a circular velocity of 90~km~s$^{-1}$ -- marked with corresponding purple squares.}
	\label{fig:models_v}
\end{figure*}

\begin{figure}
	\includegraphics[width=0.97\columnwidth]{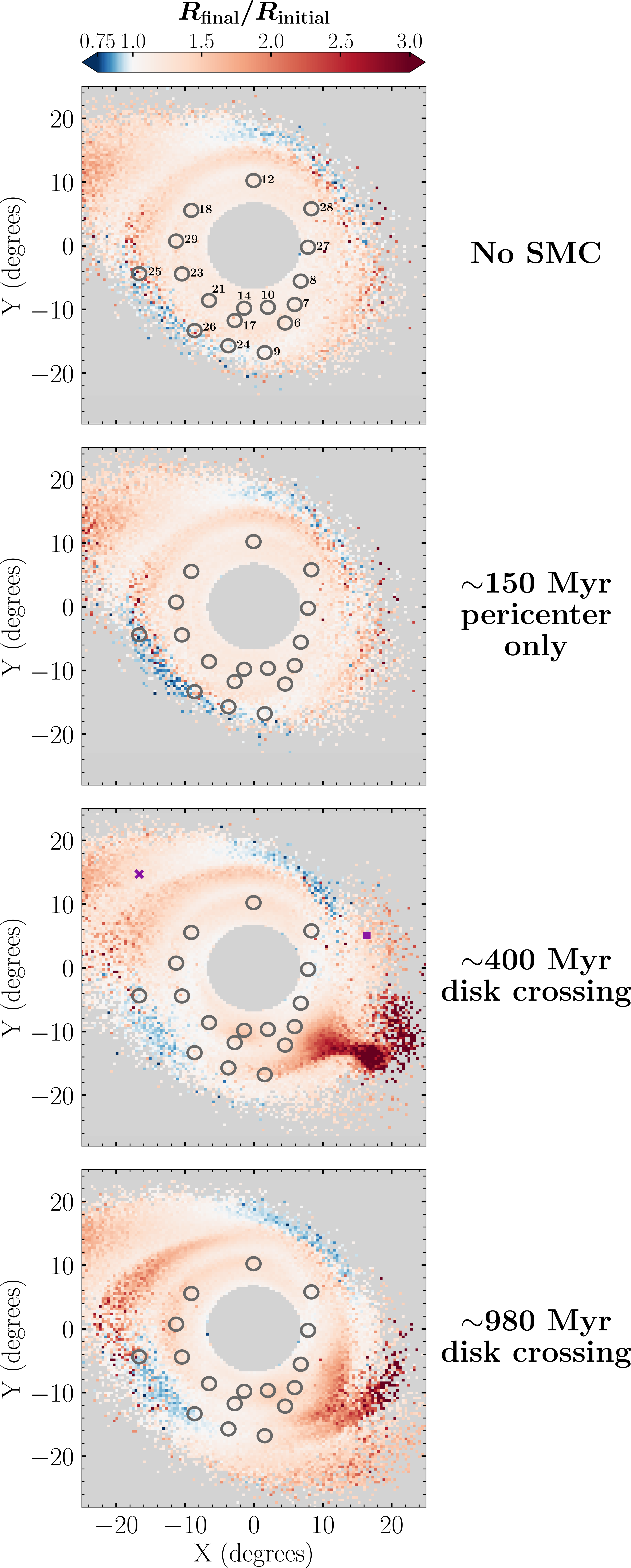}		
	\caption{
			Ratio of final to initial particle in-plane galactocentric radius ($R_{\text{final}}/R_{\text{initial}}$) for the same four individual model realisations as in Fig.~\ref{fig:models_v}. Un-filled grey circles represent the locations of the MagES fields studied in this analysis; the top panel indicates field numbers, for reference. The centre 7$^\circ$ of each model is masked to emphasise the kinematics of the outskirts. In the third row, the location of the \textasciitilde400~Myr crossing, at the crossing time, is marked by a purple x-sign, with the approximate present-day location -- computed by rotating the location of the original crossing within the LMC’s disk plane, assuming a circular velocity of 90~km~s$^{-1}$ -- marked with a corresponding purple square.	}
	\label{fig:models_r}
\end{figure}

In order to show the qualitative effects of these varying interaction histories on the LMC, Figs.~\ref{fig:models_v} and \ref{fig:models_r} present current-day kinematic maps for four individual model realisations. Each realisation is selected to have a similar orbit of the LMC around the MW (and thus similar effects from MW tides), but varies in its interactions with the SMC. The top row presents a “no-SMC” realisation, with the second row presenting the same realisation (i.e. that which has identical present-day systemic position and velocity for the LMC) in the base-case model as the top row. This model experiences only the most recent SMC pericentre \textasciitilde150~Myr ago. The third row presents a base-case realisation which experiences the \textasciitilde400~Myr SMC crossing of the LMC disk plane in addition to the most recent SMC pericentric passage, whilst the last row presents one of the few base-case realisations which experiences an SMC crossing of the LMC disk plane, \textasciitilde980~Myr ago in addition to the \textasciitilde400~Myr disk crossing and \textasciitilde150~Myr pericentric passage. 

For each realisation, we show the mean within each spatial bin of four key current-day kinematic and structural properties of the models. Fig.~\ref{fig:models_v} presents, in order: i) the in-plane radial velocity ($V_r$), ii) the out-of-plane vertical velocity ($V_z$), and iii) the out-of-plane distance ($z$) in each column of the figure, while Fig.~\ref{fig:models_r} presents the ratio of final to initial particle in-plane radius ($R_{\text{final}}/R_{\text{initial}}$). We do not show maps for the azimuthal velocity ($V_\theta$) as in our models, this quantity shows an inherent systematic decrease with radius beyond 10~kpc (where we match the observed rotation curve of the LMC) reflecting the Hernquist potential used to model the LMC halo. As a consequence, in our models this velocity component is more difficult to interpret in the extreme LMC outskirts, and thus less useful as a tracer of interaction, compared to the radial and vertical motions.

In order to facilitate qualitative comparison of the models with our observations, for each realisation we show the $X,Y$ location of the MagES fields discussed in this paper using grey circles. In the case of Fig.~\ref{fig:models_v}, the fill colour of the circles is set according to the observed kinematics of each field, using the same scale as applied to the models. While this means that model spatial bins which lie within the region covered by each field are not visible for a direct comparison to the observed kinematics, we note that model kinematics do not change significantly over the \textasciitilde2-degree spatial scale of the MagES fields. As such, comparison of the observed kinematics to the surrounding spatial bins of the model provides a reasonable indication of qualitative similarities or differences between the observed and model kinematics in that region. In Fig.~\ref{fig:models_r}, since $R_{\text{initial}}$ cannot be determined from our data, the field circles, while plotted, remain unfilled (i.e. model values of $R_{\text{final}}/R_{\text{initial}}$ are visible throughout).

Note that our aim in also presenting MagES data in Fig.~\ref{fig:models_v} is to facilitate qualitative comparisons between the models and our data for each of the disk regions and substructures discussed in \S\ref{sec:northeast}-\ref{sec:southarm}, and not necessarily the LMC as a whole. For example, if a given model realisation displays qualitatively similar kinematics\footnote{i.e., having the same direction and a similar order of magnitude as those observed; as indicated by having a similar, though not necessarily precisely matching, colour between the MagES field and the surrounding model.} to those observed for a specific disk region in each of the three columns of Fig.~\ref{fig:models_v}, this suggests the interactions experienced by that model realisation can potentially play a role in producing the (disturbed) kinematics observed in that region. We discuss these comparisons in further detail in Section~\ref{sec:comp}. 

\subsection{Predicted effects of interactions in the outer LMC}\label{sec:modpred}
Figs.~\ref{fig:models_v} and \ref{fig:models_r} reveal the broad kinematic trends we expect to result from different perturbations to the LMC disk. We first consider the effect of the MW tides on the LMC disk, shown in the top row of each figure. The MW’s most significant effect on the LMC is production of a strong vertical velocity gradient across the LMC disk, with a strongly positive $V_z$ in the far northeast, and a strongly negative vertical velocity in the far southwest. This is consistent with our findings in \citetalias{C21} of an increasingly positive $V_z$ along a long arm-like feature to the north of the LMC. However, we note at the distances of the disk fields analysed in this paper (as opposed to those along the long ‘arm-like’ southern substructure), the magnitude of the $V_z$ perturbation in the model is generally small (for these models, on the order of \textasciitilde10~km~s$^{-1}$). 

The MW additionally has a (somewhat milder overall) effect on the in-plane radial velocity, acting to compress the LMC disk in the northwest and southeast (marked by both a negative $V_r$ and $R_{\text{final}}/R_{\text{initial}}$<1), and mildly elongate it in the northeast and southwest (marked by $R_{\text{final}}/R_{\text{initial}}$>1). Like the vertical perturbation, these effects are strongest at large LMC galactocentric radii, particularly near the locus of the long ‘arm-like’ southern substructure, and comparatively weak (with the magnitude of $V_r$ on the order of \textasciitilde10~km~s$^{-1}$, and $0.75{\lesssim}R_{\text{final}}/R_{\text{initial}}{\lesssim}1.25$ for these models) at the radii of the MagES disk fields. We do note that whilst the magnitude of this effect is not as dramatic as that in $V_z$, some level of compression (i.e. $R_{\text{final}}/R_{\text{initial}}$<1) in the southeast and northwest LMC outskirts is maintained across all model realisations shown. 

Comparing the MW-only model (top row of Figs.~\ref{fig:models_v}-\ref{fig:models_r}) to the model which includes only the SMC’s most recent pericentre (second row of Figs.~\ref{fig:models_v}-\ref{fig:models_r}), we see these are not very significantly different. The south-eastern edge of the LMC is more strongly compressed (with $R_{\text{final}}/R_{\text{initial}}$ mildly smaller than in the MW-only case), though the predicted radial velocity does not change significantly in this region. In addition, there are localised regions in the southwest of the LMC -- in the direction of the projected location of the SMC’s closest pericentric distance -- that have more strongly negative vertical velocities and/or more strongly positive radial velocities than in the MW-only model. These localised perturbations match the direction of the SMC’s gravitational force on the LMC during its recent pericentric passage. However, because the pericentric passage is at a moderate out-of-plane distance (\textasciitilde$-$9~kpc for this particular model realisation)\footnote{While this is on the higher end of the $-6.8^{+2.5}_{-2.6}$~kpc pericentric distance found across the base-case model suite, we find the kinematic signatures in this realisation are not significantly different than those models which experience closer pericentric passages, albeit at a strength on the lower end of those observed.}, the kinematic signatures of this perturbation are not as strong as interactions which impact the LMC’s disk plane directly, although we note the lack of self-gravity in our models may mean these effects are mildly underestimated (see \citetalias{C21}). 

More dramatic effects on the kinematics of the disk are produced by SMC crossings of the LMC disk plane, as is evident comparing the second row of Figs.~\ref{fig:models_v}-\ref{fig:models_r} -- which include only the most recent pericentric passage -- to the third row of Figs.~\ref{fig:models_v}-\ref{fig:models_r}, which display a model realisation where the SMC additionally crosses the LMC disk plane \textasciitilde400~Myr ago. In this particular model realisation, the SMC crosses the LMC disk plane in the then-northeast of the LMC, at a radial distance of 17.7~kpc (marked by a purple cross in the figures). The material closely perturbed by this interaction has since moved clockwise with the LMC’s rotation and is now located in the western outskirts of the LMC. 

This is evident in the strongly perturbed kinematics of this region in our model, and is also seen in panel d of Fig.~14 of \citetalias{C21} (which shows the same model realisation as in the third row of Figs.~\ref{fig:models_v}-\ref{fig:models_r}, and colour-codes the present-day particle distribution by the distance of each particle from the SMC at the time of the disk plane crossing). The in-plane radial velocity is strongly positive ($\geq$50~km~s$^{-1}$) across the western LMC disk outskirts, and $R_{\text{final}}/R_{\text{initial}}$ is substantially greater than one in the southwest. As the SMC crosses the disk plane at a (radial) distance of 17.7~kpc in this particular realisation, these effects are consistent with the direction of the SMC’s gravitational force on the LMC in this region during the crossing. We note the radial distance of the crossing in this particular realisation is at the lower end of the distribution of radial crossing distances for this event seen in our models\footnote{The median radial crossing distance is 28.8~kpc with 1$\sigma$ thresholds of 39.8 and 19.6~kpc.}, which means the gravitational force of the SMC (and thus the resultant kinematic signatures in the LMC disk) are on the stronger end of those allowable\footnote{Analysis of all base-case model realisations which experience the \textasciitilde400Myr crossing reveals qualitatively similar kinematic trends to those shown in Figs.~\ref{fig:models_v}-\ref{fig:models_r} are observed for models which have crossing distances less than the median \textasciitilde29~kpc, with more distant crossings having comparatively negligible impact on the kinematics of the LMC disk.}. 

In terms of vertical perturbations in this model realisation, we see a strongly positive vertical velocity ($\geq$40~km~s$^{-1}$) and out-of-plane distance (>4~kpc) in the south-western outskirts of the disk, and a strongly negative out-of-plane distance ($\leq-$4~kpc) and vertical velocity ($-25$~km~s$^{-1}$) along the rest of the western edge of the LMC. However, we do note the vertical velocity in the north-western region is not significantly more negative than that introduced by the Milky Way. The asymmetry in out-of-plane distances, and to some extent vertical velocities (i.e. positive in the southwest, and negative in the northwest) is expected from the SMC crossing the disk in a direction from positive to negative $z$. This is because material leading the crossing point relative to the sense of (clockwise) rotation in the LMC -- today in the southwest -- feels a net positive vertical perturbation from the SMC, and material trailing the crossing point -- today in the northwest -- feels a net negative vertical perturbation from the SMC. 

In addition to the \textasciitilde400~Myr disk crossing, which is experienced by \textasciitilde50\% of our model realisations, some model realisations also experience a SMC crossing of the LMC disk plane \textasciitilde1~Gyr ago. In these cases, the SMC crosses the disk plane in the opposite sense (i.e., from negative to positive $z$) to that \textasciitilde400~Myr ago. The particular realisation shown in the bottom row of Figs.~\ref{fig:models_v}-\ref{fig:models_r} experiences this additional disk crossing \textasciitilde980~Myr ago, in the then-southwest of the LMC at a radial distance of 8.2~kpc. This is a distance \textasciitilde40\% smaller than the disk crossing \textasciitilde400~Myr ago. We thus expect even stronger perturbative effects from the SMC during this crossing. 

However, the orbital period of material around the LMC disk is longer at larger radii (due to the approximately constant circular velocity). At the crossing radius, the time since the crossing is comparable to the orbital time, and material closely perturbed in this interaction is in approximately the same location as when the perturbation occurred (i.e. the south-western LMC). In contrast, the shorter orbital period at smaller radii is such that material can complete more than one full disk rotation in the time since the crossing, so material closely perturbed by the interaction appears somewhat mixed (and is even present in the current north-east of the LMC: see panel f of Fig.~14 in \citetalias{C21}). Consequently, the precise kinematic signatures of this earlier perturbation are more difficult to disentangle, with this made even harder by the fact that some material closely affected by this interaction is subsequently perturbed during the SMC crossing of the disk plane \textasciitilde400~Myr ago. However, we do still note a more strongly negative in-plane radial velocity and out-of-plane vertical velocity, as well as a mildly more positive out-of-plane distance, in the extreme southern outskirts of the LMC in this model realisation as compared to that in the third row of Figs.~\ref{fig:models_v}-\ref{fig:models_r}, which are likely the result of this older disk crossing.  

\subsection{Comparing models to observations}\label{sec:comp} 
We now perform a qualitative comparison of our measured kinematics to Figs.~\ref{fig:models_v} to learn more about the potential origins of the various kinematic signatures observed across the LMC outskirts. We first note that kinematics in the vicinity of the three fields in the north-eastern LMC disk are only very mildly perturbed in each of the model realisations. Even in the most disturbed realisation, which experiences disk crossings at \textasciitilde980 and \textasciitilde400~Myr in addition to the SMC’s most recent pericentric passage (bottom row of Fig.~\ref{fig:models_v}), the radial and vertical velocities near fields 18 and 29 (located at $X,Y$ positions $-$8.4,5.8 and $-$10.7,1.0 respectively) are near zero, as are the out-of-plane distances. This is consistent with our observations, and supports our use of data in this region to fit the LMC disk plane.  

Considering next the fields along the western edge of the LMC disk, we find the kinematics in these fields are mostly qualitatively consistent with those resulting from the SMC’s crossing of the disk plane \textasciitilde400~Myr ago (third row of Fig.~\ref{fig:models_v}), with some effect from the LMC’s infall to the Milky Way potential (top row of Fig.~\ref{fig:models_v}). Fields 27 and 8 (located at $X,Y$ positions of 8.4,$-$0.5 and 7.2,$-$5.8 respectively) have positive in-plane radial velocities, strongest in field 8 in the south-western disk. This is qualitatively consistent with the strongly positive radial velocity produced by the disk crossing in the western LMC outskirts, though we note the strongest model $V_r$ signature in the model is located at greater radii than the observed disk fields. The near-zero radial velocity observed in field 17 (located at $X,Y$ position $-$2.5,$-$11.7) can be interpreted as resulting from the combined effect of the disk crossing, which pulls material radially outward, and the compressive force of the MW tides, which pushes material radially inward in the north-western disk.

Fields 8 and 27 are also qualitatively consistent with the vertical velocity asymmetry resulting from the \textasciitilde400~Myr crossing (i.e. positive $V_z$ in the southwest and negative $V_z$ in the northwest) if field 27 is located near, or in line with\footnote{As the rotation curve of the LMC is approximately flat, material at a smaller LMC galactocentric radius is able to rotate through a larger range of position angles in a given period of time than material at a greater galactocentric radius. Consequently, the position angle of material “in line with” the current-day location of the disk crossing varies with galactocentric radius. As position angles are measured from north towards east in this paper, material at smaller galactocentric radii than the current-day location of the disk crossing is located at numerically smaller position angles than the disk crossing location, while material at larger galactocentric radii is located at numerically larger position angles than the disk crossing location.}, the current-day position of the disk crossing point itself (where $V_z$\textasciitilde0). In Figs.~\ref{fig:models_v}-\ref{fig:models_r}, we see the the current-day location of the crossing point for the particular realisation displayed in the third row of the figure -- marked by a purple square -- is broadly in line with field 27. In addition, although not shown in Fig.~\ref{fig:models_v}, we find the velocity dispersions in the western LMC disk outskirts are mildly elevated in model realisations experiencing the \textasciitilde400~Myr crossing compared to those which only experience the SMC’s most recent pericentric passage. This is consistent with the elevated velocity dispersions observed in these fields.

However, we do note model realisations which experience only the \textasciitilde400~Myr disk crossing (as in the third row of Fig.~\ref{fig:models_v}) predict the vertical velocity and out-of-plane distance in the vicinity of field 28 should be negative. This is inconsistent with the positive values observed. We speculate these may reflect the effects of earlier interactions with the SMC. We note the model realisation in the bottom row of Fig.~\ref{fig:models_v} (which additionally experiences a disk crossing \textasciitilde980~Myr ago) has a positive vertical velocity and out-of-plane distance extending to larger position angles -- i.e. closer to field 28 -- than that in the third row of Fig.~\ref{fig:models_v} (which experiences only the \textasciitilde400~Myr crossing). However, as the location and timing of older interactions with the SMC are poorly constrained in our simple models, more detailed modelling is required to confirm this.  

We next discuss fields in the southern LMC disk. The increasing in-plane radial velocities with position angle observed in this region can be understood as a combination of forces. In the southeast, the compressive forces from the Milky Way tides produce negative radial velocities, as observed in field 23 (located at $X,Y$= $-$10.0,$-$4.2). Moving toward the southwest, the gravitational force of the SMC becomes increasingly dominant, pulling material outward and producing positive radial velocities as observed in fields 10 and 8 (located at $X,Y$ positions of 2.3,$-$9.8 and 7.2,$-$5.8 respectively). In this south-western region, both the \textasciitilde400~Myr disk crossing (third row of Fig.~\ref{fig:models_v}) and the SMC’s recent pericentric passage (second row of Fig.~\ref{fig:models_v}) can produce positive radial velocities. 

These southern disk fields are also observed to transition from negative to positive vertical velocities moving around the disk from southeast to southwest. This is inconsistent with model realisations which experience only the recent SMC pericentre (second row of Fig.~\ref{fig:models_v}), which have negative vertical velocities throughout the southern outskirts, and model realisations which experience only the \textasciitilde400~Myr disk crossing (third row of Fig.~\ref{fig:models_v}), which have positive vertical velocities throughout the southern outskirts. The transition is in fact best matched by the model realisation which experiences SMC crossings of the LMC disk plane at both \textasciitilde400~Myr and \textasciitilde980~Myr ago (bottom row of Fig.~\ref{fig:models_v}). The increased vertical velocity dispersion and (positive) out-of-plane distance observed in field 17 (located at an $X,Y$ position of $-$2.5,$-$11.7) relative to field 14 (located at an $X,Y$ position of $-$1.1,$-$9.8) additionally suggest this region is affected by a perturbation strongest in the $z$-direction -- i.e. an SMC crossing of the LMC disk plane.  However, more detailed models will be required to confirm the influence of older interactions, and how these combine with the effects of more recent interactions, in this region. 

While our simple models can generally provide reasonable descriptions of the kinematics of fields in the outer LMC disk, they are not able to fully replicate the kinematics of individual substructures in the southern regions. The mildly positive in-plane radial velocities and elevated radial velocity dispersions in fields 6 and 7 (located in the claw-like Substructure 2, at $X,Y$ coordinates 4.8,$-$12.3 and 6.2,$-$9.4 respectively) are similar to those in the nearby southern disk. This is consistent with an origin in recent interactions with the SMC, be that the \textasciitilde400~Myr disk-plane crossing (third row of Fig.~\ref{fig:models_v}) or the pericentric passage \textasciitilde150~Myr ago (second row of Fig.~\ref{fig:models_v}). Yet none of our models can replicate their very strongly perturbed azimuthal and vertical velocities, or the distinctive shape of the feature. We thus speculate this substructure is potentially the result of repeated interactions with the SMC beyond those occurring in the 1~Gyr for which our model suites are run, each having an additive effect. Such a scenario could plausibly produce the large velocity perturbations observed in these fields. However, self-gravitating models will be required to confirm this, as it is possible the lack of self-gravity in our models simply underestimates the effect of recent interactions with the SMC in this region. 

Our models are also unable to replicate the effectively unperturbed kinematics of field 26 in the “arm-like” southern substructure (at an $X,Y$ location of $-$8.4,$-$13.1). Each of our models predicts a strongly negative radial velocity at this location, and also significantly underestimates the out-of-plane distance ($z$) -- though this is perturbed in the correct direction. The large out-of-plane distances of fields 9 and 24 (located at $X,Y$ coordinates of 1.7,$-$16.9 and $-$3.6,$-$15.6 respectively) are similarly significantly underestimated in our simple models. However, we do note that the negative vertical and radial velocities in these fields are qualitatively similar to the model realisation which experiences SMC crossings of the LMC disk plane at both \textasciitilde980 and \textasciitilde400~Myr ago (bottom row of Fig.~\ref{fig:models_v}). More detailed modelling is clearly required to understand the origin of these velocity perturbations. 

\subsection{Effect of disk geometry}\label{sec:inc_discuss}
As mentioned in Section~\ref{sec:models}, the inclination of the LMC disk calculated in this paper ($36.5^\circ\pm0.8^\circ$) is somewhat higher than that of the dynamical models we subsequently compare to \citep[which have an inclination $i$\textasciitilde$25^\circ$, taken from][]{choiSMASHingLMCTidally2018}. These models were initially designed to investigate the LMC's northern arm feature in \citetalias{C21}, and thus computed prior to our current analysis. It is beyond the scope of this paper to recompute another full suite of dynamical models, with varying LMC/SMC/MW masses, utilising an updated disk geometry consistent with that derived in this work. However, we do compute a smaller subset of models, otherwise identical to the “base-case” models but using a \textasciitilde10$^\circ$ higher disk inclination, for a first brief investigation into the potential effect of the disk orientation on our results.

We find a similar range of LMC-SMC interactions occur in the high-inclination models, including realisations which experience only the SMC's most recent pericentric passage, as well as model realisations which experience SMC crossings of the LMC disk plane at \textasciitilde400~Myr and \textasciitilde1~Gyr ago. The frequency and location of these interactions do vary between the low- and high-inclination models. This is due to the difference in the orientation of the LMC potential relative to the SMC, which has a small but observable effect on the SMC's orbit. For example, some model realisations which experience a SMC disk crossing \textasciitilde400~Myr ago at a very large galactocentric radius in the low-inclination models, do not experience this crossing in the higher-inclination model realisations. In such realisations, the crossing only just grazes the disk plane in the low-inclination model, with the changed inclination of the LMC disk plane in the higher-inclination model subset resulting in the SMC not formally crossing the disk plane in this case. 
	
In future work, we aim to more completely explore the orbital parameter space of the Clouds, including variation of the LMC geometry across the range of values measured in the literature, in order to more precisely characterise the frequency and properties of different LMC-SMC interactions. For the moment, however, the subset of models computed with the higher inclination indicate that in general, when a given type of interaction (i.e. pericentric passage or disk plane crossing) occurs, the resulting kinematic differences appear very similar irrespective of the assumed geometry. Structural and kinematic maps for high-inclination realisations show qualitatively similar trends to those in Figs.~\ref{fig:models_v}-\ref{fig:models_r}, and do not change our conclusions regarding which regions of the LMC are most significantly impacted by different possible interactions.

\section{Summary and Conclusions}\label{sec:concs}
In this paper, we have explored the structural and kinematic properties of the LMC outer disk and surrounding substructures. Our analysis utilises spectroscopic data for red clump and red giant branch stars across 18 MagES fields to obtain [Fe/H] abundances, and in conjunction with Gaia EDR3 astrometry, 3D kinematics for stars across the LMC outskirts. Ten MagES fields probe the outer LMC disk at galactocentric radii of 8.5$^\circ$<$R$<11$^\circ$, five trace the long “arm-like” southern substructure discovered in \citet{belokurovCloudsArms2019}, and three are located on the claw-like southern substructures discovered in \citet{mackeySubstructuresTidalDistortions2018}. We also use Gaia EDR3 photometry of Magellanic red clump stars to probe the structure of the LMC outskirts.

We find field 3, located near the western end of the southern “arm-like” structure, is comprised predominantly of SMC, not LMC, material, owing to its distinct SMC-like kinematics and comparatively low ([Fe/H]\textasciitilde$-1.4$) metallicity. At a SMC galactocentric radius of 9.5$^\circ$, it represents one of the most distant detections of SMC debris. This field is under further investigation by the MagES team, and is not considered further in this paper. The other 17 fields studied each have properties more consistent with the LMC, and are thus analysed in this context. 

We find an approximately constant mean metallicity of [Fe/H]\textasciitilde$-$1 across all considered fields, consistent with previous literature measurements in the outer LMC disk. Where available, [Fe/H] measurements for individual RGB stars reveal an \textasciitilde0.5~dex standard deviation in each field. RC photometry for each field also shows consistent $(G_{BP}-G_{RP})_0$ colours and colour dispersions across all fields. In combination with the consistent metallicity measurements, this indicates similar stellar populations are present within each field, and that Magellanic stars in all fields are predominantly LMC disk material. 

Motivated by previous findings that the north-eastern outskirts of the LMC disk are relatively kinematically undisturbed \citepalias{C20}, we utilise Gaia EDR3 astrometry and photometry for RC stars in the north-eastern outskirts of the LMC disk (9.5$^\circ$<$R$<10.5$^\circ$, 5$^\circ$<$\Phi$<90$^\circ$) to fit an inclined disk model as described in \citet{vandermarelNewUnderstandingLarge2002} to the LMC. In addition to describing the geometry of the disk and the azimuthal velocity of the LMC, our model also outputs a “reference magnitude” which describes the apparent brightness of a red clump, comprised of an identical stellar population to that of our fields in the outskirts, located at the centre distance \citep[49.59~kpc:][]{pietrzynskiDistanceLargeMagellanic2019} of the LMC.

We find at these large galactocentric radii, the LMC disk has an inclination of $36.5^\circ\pm0.8^\circ$, a position angle of the LON of $145.0^\circ\pm2.5^\circ$, and an azimuthal velocity of $69.9\pm1.7$~km~s$^{-1}$. These are consistent with several literature measurements of the LMC disk at smaller radii. Our “reference magnitude” for the red clump is $G_0$\textasciitilde18.9, and we utilise this, in conjunction with RC photometry for each of the analysed fields, to derive absolute distance estimates across the LMC outskirts and convert our observed field kinematics into the frame of the LMC disk. 

Our results are qualitatively compared to a suite of simple dynamical models of the Magellanic system described in \citepalias{C21}, sampling from uncertainties in the central locations and systemic motions of the LMC and SMC. This allows investigation of how different interactions between the LMC, SMC, and MW can potentially produce the observed perturbations. Four model realisations are considered, each with similar orbits of the LMC around the MW. The first omits the SMC entirely, the second experiences only the most recent SMC pericentric passage around the LMC \textasciitilde~150~Myr ago. The third and fourth realisations experience SMC crossings of the LMC disk plane in addition to the most recent pericentric passage, with the third realisation having an SMC crossing of the LMC disk plane \textasciitilde400~Myr ago, and the fourth having disk plane crossings both \textasciitilde400 and \textasciitilde980~Myr ago. Findings for each kinematically distinct region of the LMC are summarised below.

As expected, fields in the north-eastern LMC disk display relatively unperturbed kinematics, with radial and vertical velocities consistent with zero, and only minor deviations in a field near base of an arm-like feature north of the LMC. We note that even in the most disturbed model (i.e. that which experiences disk crossings \textasciitilde400 and \textasciitilde980~Myr ago, as well as the pericentric passage \textasciitilde150~Myr ago) the north-eastern LMC has relatively undisturbed kinematics at the locations of the MagES fields, consistent with our observations.

We find the western outskirts of the LMC are most strongly perturbed in an SMC crossing of the LMC disk plane \textasciitilde400~Myr ago, with the observed kinematics of fields in the western disk mostly qualitatively similar to those predicted from this interaction. The in-plane radial velocity uniformly decreases moving northwards along the disk edge, which we find is consistent with the combined effects of the MW tides acting to compress material in the north-western LMC, and the effect of the SMC during the disk crossing, which acts to pull material radially outward along the entire western LMC edge. Our models additionally predict that disk material leading the crossing point during the \textasciitilde400~Myr crossing (today in the southwestern LMC) feels a net positive vertical perturbation -- consistent with our observations in the southwestern disk -- but that material trailing the crossing point (today in the northwest) feels a net negative vertical perturbation. This is inconsistent with our observations in the northwest. We speculate the north-western LMC may have been further perturbed by older interactions with the SMC. However, as such interactions are relatively poorly constrained in our simple models, further investigation is required. 

Fields in the southern LMC disk also likely feel perturbative effects from multiple interactions with the SMC, with velocity dispersions in these fields approximately double those in the unperturbed north-eastern disk. Clear kinematic trends are observed moving westward around the disk, with both the in-plane radial velocity and out-of-plane vertical velocity increasing from negative to positive values, and the out-of-plane distance also increasing. The in-plane radial velocity trend is consistent with the combined effects of MW tides, which compress material in the southeast, and recent interactions with the SMC, with both the recent pericentric passage and \textasciitilde400~Myr disk crossing pulling material in the southwest radially outward. However, the increasing vertical velocity is best replicated by model realisations which experiences disk crossings at both \textasciitilde400 and \textasciitilde980~Myr ago. We additionally find field 17, located in the claw-like Substructure 1, has very similar kinematics to those in the nearby southern disk. This indicates it experiences similar perturbative effects as the nearby southern disk. 

In contrast to the LMC disk, the significant perturbations observed in individual substructures in the LMC outskirts are not well described by our models. Fields located on the claw-like Substructure 2 are significantly perturbed in the $z$ direction, with highly elevated vertical velocities (and dispersions), and out-of-plane distances on the order of \textasciitilde8~kpc. Additionally, the feature appears to be counter-rotating relative to the LMC disk, with $V_\theta$\textasciitilde$-$20~km~s$^{-1}$. Our models cannot reproduce the observed kinematics, nor the shape of the feature, and we speculate that additional interactions with the SMC, prior to the 1~Gyr for which our models are run, are potentially required to produce this feature.

Finally, we discuss the “arm-like” southern substructure, which we find does not display coherent kinematic trends along its length. Fields toward the eastern end of the feature are not too significantly (<2$\sigma$) disturbed from the expected disk kinematics, with field 26 appearing almost undisturbed if not for a \textasciitilde6~kpc out-of-plane distance. However, our models suggest these fields should be more significantly disturbed, particularly in terms of their in-plane radial velocity, than is observed. In contrast, fields further west in the substructure do display more disturbed kinematics. While these fields have negative radial and vertical velocities, qualitatively similar and in the correct direction for the feature to be a counterpart to the northern arm-like feature in \citetalias{C21}, these fields are comparatively much more kinematically hot (with velocity dispersions up to \textasciitilde40~km~s$^{-1}$) than the northern arm. A qualitative match to the velocities in these fields is given by model realisations which experience SMC crossings of the LMC disk plane \textasciitilde980 and \textasciitilde400~Myr ago, though the out-of-plane distances (>10~kpc) of these fields are significantly underestimated. 

In conclusion, the simple models presented here, in conjunction with our extensive observations, provide a useful first exploration of how interactions in the Magellanic system can potentially produce perturbed structural and kinematic properties in the LMC outskirts. It is clear from our analysis that different regions and substructures in the outskirts of the Clouds are sensitive to different events in the Magellanic interaction history, such that a joint analysis of each of the observed features can be used to understand the overall interaction history of the Clouds. In particular, our observations of the substructures in the southern outskirts of the LMC, in combination with future self-gravitating models that are able to more accurately trace the dynamical influence of the SMC over longer timescales, will be critical in placing tight constraints on the early orbital history of the Magellanic Clouds, and the consequences for their star formation histories.

\section*{Acknowledgements}

This work has made use of data from the European Space Agency (ESA) mission {\it \textit{Gaia}} (\url{https://www.cosmos.esa.int/gaia}), processed by the {\it \textit{Gaia}} Data Processing and Analysis Consortium (DPAC, \url{https://www.cosmos.esa.int/web/gaia/dpac/consortium}). Funding for the DPAC has been provided by national institutions, in particular the institutions participating in the {\textit{Gaia}} Multilateral Agreement. Based on data acquired at the Anglo-Australian Observatory. We acknowledge the traditional owners of the land on which the AAT stands, the Gamilaraay people, and pay our respects to elders past, present and emerging. This research has been supported in part by the Australian Research Council (ARC) Discovery Projects grant DP150103294. ADM is supported by an ARC Future Fellowship (FT160100206).

\section*{Data availability}
The data underlying this article will be shared on reasonable request to the corresponding author.




\bibliographystyle{mnras}
\bibliography{p3refs} 



\appendix

\section{LMC geometry model fitting}\label{sec:appendixeq}
\begin{figure*}
	\includegraphics[width=\textwidth]{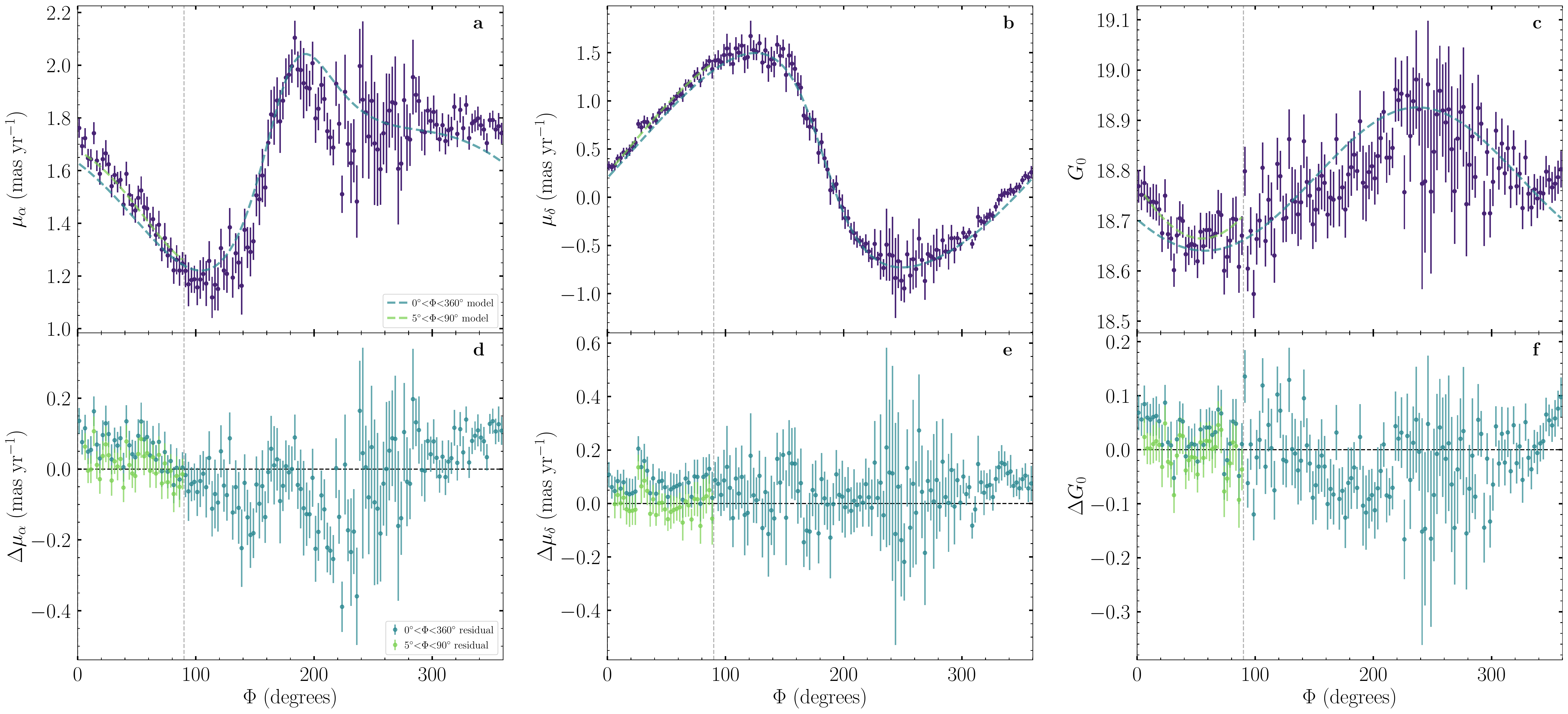}
	\caption{Upper panels: proper motions (left, centre) and $G_0$ magnitude (right) as a function of position angle for the LMC disk. Purple points represent bin median values for Magellanic red clump stars at LMC galactocentric radii between 9.5$^\circ$ and 10.5$^\circ$ used to fit the LMC disk plane, selected from Gaia EDR3 as described in Section~\ref{sec:fitdisk}. Error bars represent the standard deviation within each bin; large error bars for bins in the western outskirts of the disk ($240^\circ\lesssim\Phi\lesssim280^\circ$) are due to comparatively low numbers of stars per bin compared to the remainder of the disk, as expected from Fig.~\ref{fig:map}. Coloured dashed lines show the best-fitting model predictions for the entire disk (blue) and just the northeastern disk ($5^\circ$<$\Phi$<$90^\circ$: green). Grey dashed lines are located at $\Phi{=}90^\circ$. Lower panels present residuals (data -- model) for each component, with point colours indicating the associated model.}
	\label{fig:resid}
\end{figure*}

In this appendix, we discuss the fitted model described in Section~\ref{sec:fitdisk} to describe the geometry of the LMC disk. Fig.~\ref{fig:resid} shows the fit results. Panels show, in order from left to right, the mean $\mu_\alpha$,$\mu_\delta$, and $G_0$ magnitude in each bin, as a function of position angle ($\Phi$) around the LMC disk. Purple points in the upper panels show the observed data used to perform the fits. Dashed lines in the upper panels show  $\left\{\mu_{\alpha,\text{mod}},\mu_{\delta,\text{mod}},m_{\text{mod}},\right\}$ as calculated using the best-fitting model values for $\left\{i,\Omega,V_\theta,G_0^{\text{RC}}\right\}$ when (blue) the full range of position angles are fitted, and (green) only bins between $5^\circ{<}\Phi{<}90^\circ$ are fitted. Lower panels show residuals (calculated as $x_{\text{obs}}-x_{\text{mod}}$, where $x$ is, in order, $\mu_\alpha,\mu_\delta,m$) for each bin, with point colours indicating the model used in the calculation. 

When fitting our model to the full range of position angles (blue lines/points in Fig.~\ref{fig:resid}), we find there are regions with significant (i.e. non-zero) residuals, particularly in the range $-30^\circ{<}\Phi{<}80^\circ$ covering the northeastern LMC disk. This is significantly different from findings in \citetalias{C20} and \citetalias{C21} that this region of the disk is relatively undisturbed from equilibrium disk kinematics (and thus should be well-described by the simple model fitted here). Further, when fit to the full range of position angles, $V_\theta$ is poorly constrained: a value of ${<}60$~km~s$^{-1}$ (on the lower bound of those allowed in the fit) is preferred. This is considerably lower than that derived for MagES fields 12 and 18 in \citetalias{C20} using full 3D kinematics, as well as literature measurements of the LMC's rotation curve using similar stellar populations \citep[e.g.][]{gaiacollaborationGaiaEarlyData2021a, vandermarelThirdEpochMagellanicCloud2014, wanSkyMapperViewLarge2020}. This suggests our simple model is not an accurate description of the entire outer disk. 

Motivated by the results of \citetalias{C20} and \citetalias{C21}, we therefore restrict our fit to a position angle range of $5^\circ{<}\Phi{<}90^\circ$ (green lines/points in Fig.~\ref{fig:resid}). We find residuals for this fit are consistent with zero across the fitted range, and that each of the fitted parameters is well-constrained. In addition, the resulting value for $V_\theta$ is more consistent with literature measurements. We therefore consider this to be a more reliable description of the LMC geometry within this range of position angles.

\section{Proper motions across the Magellanic periphery}\label{sec:appendixpm}
\begin{figure}
	\centering
	\includegraphics[width=\columnwidth]{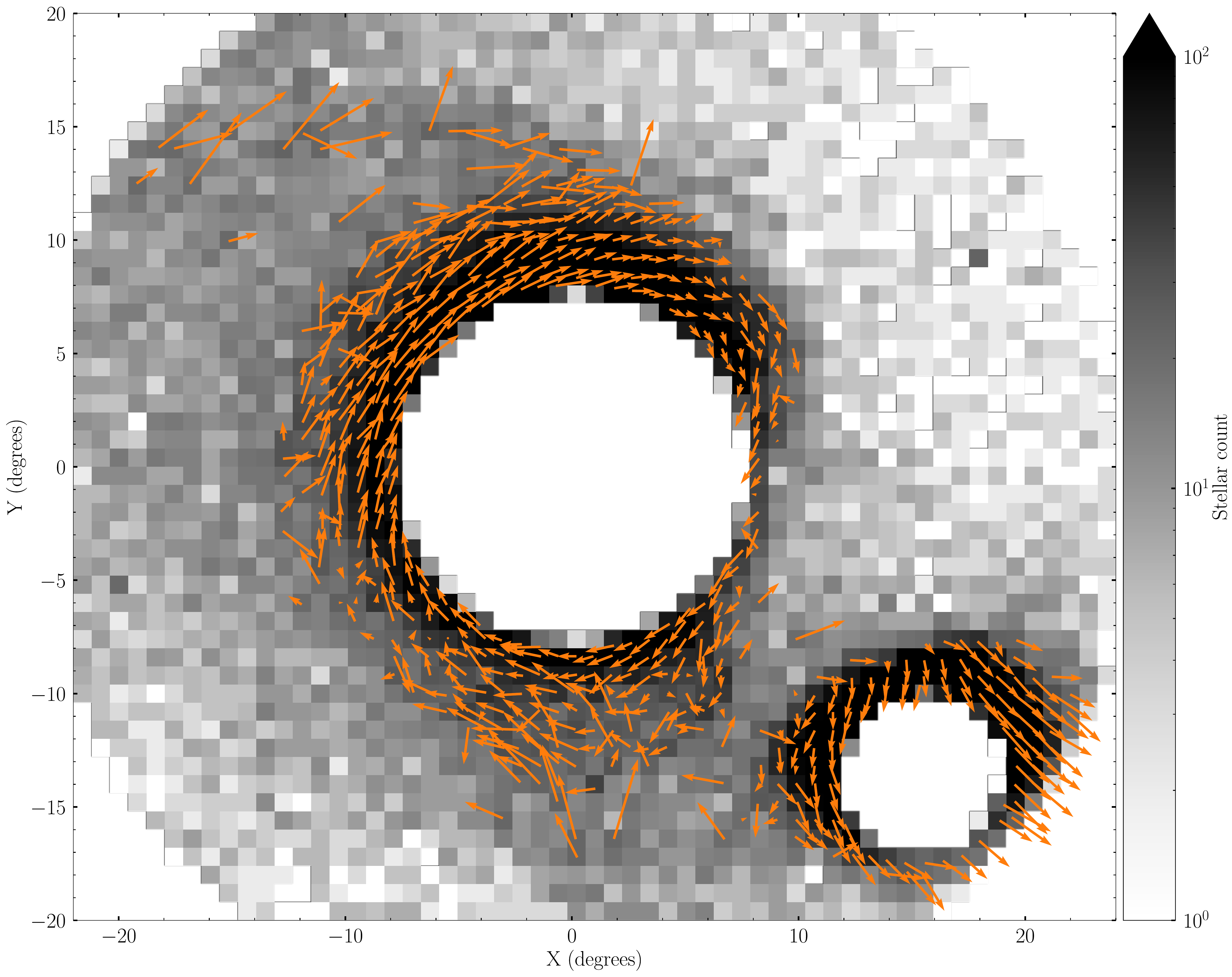}
	\caption{Proper motion vector field (orange arrows) across the Magellanic periphery, analogous to Fig.~17 of \protect\citet{gaiacollaborationGaiaEarlyData2021a}. The background image shows the stellar density of likely Magellanic stars with LMC galactocentric radii >7$^\circ$ and SMC galactocentric radii >$4^\circ$ selected from Gaia EDR3 using similar criteria to \citet{gaiacollaborationGaiaEarlyData2021a}.}
	\label{fig:pmmap_gaia}
\end{figure}

As mentioned in Section~\ref{sec:common}, it is possible to map the proper motion vector field for stars not just in MagES fields (as in Fig.~\ref{fig:vmap}), but across the entire Magellanic periphery, as in Fig.~17 of \citet{gaiacollaborationGaiaEarlyData2021a}. These maps provide a more comprehensive view of the kinematics across the entire Magellanic system than the individual MagES fields discussed in this paper. We thus present such a plot in Fig.~\ref{fig:pmmap_gaia}, using the sample of stars used to generate the density distribution in the background of Fig.~\ref{fig:vmap}. Stars in the inner LMC and SMC are omitted as these are affected by internal reddening \citep[cf. e.g.][]{choiSMASHingLMCTidally2018,tattonVMCSurveyXL2021}. 

To do so, the proper motions for each star are corrected for the LMC's bulk motion, and projected into the $x_{\text{LMC}},y_{\text{LMC}}$ coordinate system as described in Section \ref{sec:common}. Stars are subsequently binned according to their on-sky $X,Y$ position. Vectors are only plotted for bins which contain at least 20 stars, and where the net proper motion dispersion $\sigma$ within the bin (calculated as $\sigma^2 = \sigma^2_{\mu_{x_{\text{LMC}}}}+\sigma^2_{\mu_{y_{\text{LMC}}}}$) is less than 1.5~mas~yr$^{-1}$. 

It is important to note our prescription to correct for the LMC's bulk motion requires the LOS distance to each star. Unlike the MagES fields discussed in this paper -- for which distance estimates based on the red clump magnitude are available -- distances for individual stars must be assumed. For stars in the vicinity of the SMC -- defined here as having on-sky positions $X$>$7$, $Y$<$-5$, and $R_{\text{LMC}}$>$12^\circ$ -- we assume a fixed distance of 60~kpc. For all remaining stars (i.e. those in the vicinity of the LMC), we assume distances such that they lie within the plane of the LMC disk, using a disk geometry matching that derived for the northern outskirts of the LMC disk in this paper (i.e. $i=36.5^\circ$, $\Omega=145.0^\circ$). 

These assumptions are only a first approximation, and do not reflect the full range of distance perturbations in the Magellanic outskirts. As discussed in Section \ref{sec:3dkins}, much of the LMC's southern outskirts are significantly in front of the LMC disk plane; differences in the assumed distances to stars in the LMC outskirts are the primary driver of differences between Fig.~\ref{fig:pmmap_gaia} and Fig.~\ref{fig:vmap}. In addition, the SMC has a significant LOS depth, particularly in its eastern outskirts \citep[e.g.][]{hatzidimitriouStellarPopulationsLargescale1989, nideverTidallyStrippedStellar2013,tattonVMCSurveyXL2021}, which is not captured in Fig.~\ref{fig:pmmap_gaia}. Consequently, we caution over-interpretation of  Fig.~\ref{fig:pmmap_gaia}: accurate distances are necessary to fully describe the outskirts of the Clouds. 


\bsp	
\label{lastpage}
\end{document}